\documentclass[11pt]{article}
\pdfoutput=1

\newcommand\beq{\begin{eqnarray}}
\newcommand\eeq{\end{eqnarray}}
\newcommand\meff{m_{\rm eff}}
\newcommand\missET{E_T^{\rm miss}}

\usepackage{epsfig, epstopdf}

\newcommand\Zdis{Z_{\rm dis}}
\newcommand\Zexc{Z_{\rm exc}}

\newcommand\sigmab{\Delta_b}

\def\lsim{\mathrel{\rlap{\lower4pt\hbox{$\sim$}}
    \raise1pt\hbox{$<$}}}                % less than or approx. symbol
\def\gsim{\mathrel{\rlap{\lower4pt\hbox{$\sim$}}
    \raise1pt\hbox{$>$}}}            

\usepackage{graphicx}% Include figure files
\usepackage{setspace, relsize}
\usepackage{placeins}
\usepackage{multirow}
\usepackage{xcolor}
\usepackage{dcolumn}% Align table columns on decimal point
\usepackage{bm}% bold math
\usepackage{jheppub}
\usepackage{latexsym}
\usepackage{multirow}
\usepackage{color}
\usepackage{graphicx}% Include figure files
\usepackage{epsfig}  % Include figure files
\usepackage{dcolumn}% Align table columns on decimal point
\usepackage{bm}% bold math
\usepackage{dcolumn}% Align table columns on decimal point
\usepackage{textcomp}% Align table columns on decimal point
\usepackage{float}
\usepackage{hypcap}
\usepackage[]{hyperref}
\usepackage{makecell}
\usepackage{multirow}
\usepackage{color}
\usepackage{pifont}
\usepackage{subfigure}
\usepackage{amssymb}
 \usepackage{caption}
\usepackage{array}
\newcolumntype{L}[1]{>{\raggedright\let\newline\\\arraybackslash\hspace{0pt}}m{#1}}
\newcolumntype{C}[1]{>{\centering\let\newline\\\arraybackslash\hspace{0pt}}m{#1}}
\newcolumntype{R}[1]{>{\raggedleft\let\newline\\\arraybackslash\hspace{0pt}}m{#1}}

\newcommand{\wt}{\widetilde}
\newcommand{\ov}{\overline}
 \usepackage[normalem]{ulem}
\usepackage{color}

\allowdisplaybreaks
  \newcommand{\capdef}{}
  \newcommand{\mycaption}[2][\capdef]{\renewcommand{\capdef}{#2}
       \caption[#1]{{\footnotesize #2}}}
  \newcommand{\be}{\begin{equation}}
   \newcommand{\ee}{\end{equation}}

\hypersetup{
  bookmarks=true,         % show bookmarks bar?
  unicode=false,          % non-Latin characters in Acrobat?s bookmarks
  pdftoolbar=true,        % show Acrobat?s toolbar?
 pdfmenubar=true,        % show Acrobat?s menu?
 pdffitwindow=true,     % window fit to page when opened
 pdfstartview={FitH},    % fits the width of the page to the window
 pdfsubject={Neutrino Oscillations Phenomenology},   % subject of the document
 pdfnewwindow=true,      % links in new window
 pdfcreator={RevTeX},
 colorlinks=true,       % false: boxed links; true: colored links
 linkcolor=red,          % color of internal links
 citecolor=blue,        % color of links to bibliography
 filecolor=black,      % color of file links
 urlcolor=blue,           % color of external links
  }

\title{Same-sign Multilepton Signatures of an $SU(2)_R$ Quintuplet at the LHC}

\author[a,b]{Sanjib Kumar Agarwalla,}
\author[a,b]{Kirtiman Ghosh,}
\author[c]{Nilanjana Kumar,}
\author[d]{Ayon Patra}

\affiliation[a]{Institute of Physics, Sachivalaya Marg, Sainik School Post, Bhubaneswar 751005, India}
\affiliation[b]{Homi Bhabha National Institute, Training School Complex, Anushakti Nagar, Mumbai 400085, India}
\affiliation[c]{Saha Institute of Nuclear Physics, HBNI, 1/AF Bidhan Nagar, Kolkata 700064, India}
\affiliation[d]{Centre for High Energy Physics, Indian Institute of Science, Bangalore - 560012, India}

\emailAdd{sanjib@iopb.res.in}
\emailAdd{kirtiman.ghosh@iopb.res.in}
\emailAdd{nilanjana.kumar@saha.ac.in}
\emailAdd{ayon@okstate.edu}

 \abstract{We study in detail the collider signatures of an 
 $SU(2)_R$ fermionic quintuplet in the framework of 
 left-right symmetric model in the context of the 
 13 TeV LHC. Apart from giving a viable dark matter 
 candidate ($\chi^0$), this model provides unique collider 
 imprints in the form of same-sign multileptons 
through the decays of multi-charged components 
of the quintuplet. In particular, we consider the scenario 
where the quintuplet carries $(B - L) = 4$ charge, allowing for the presence of high charge-multiplicity particles with 
relatively larger mass differences among them
compared to $(B - L)$ = 0 or 2. In this paper, we mainly 
focus on the same-sign n-lepton signatures (nSSL). 
We show that with an integrated luminosity of 500 $fb^{-1}$, 
the mass of the neutral component, $M_{\chi^0} \leq 480~(800)$ 
GeV can be excluded at 95\% CL in the 2SSL (3SSL) channel 
after imposing several selection criteria. We also show that 
a $5\sigma$ discovery is also achievable if 
$M_{\chi^0} \leq 390~(750)$ GeV in the 2SSL (3SSL) 
channel with 1000 $fb^{-1}$ integrated luminosity.}

\keywords{Left-Right Symmetry, Quintuplet, LHC, Same-sign Multileptons}
\arxivnumber{1808.02904}
\begin{document}
\preprint{IP/BBSR/2018-11}
\maketitle
\flushbottom

%===============
\section{Introduction}
%===============

The Standard model (SM) of particle physics is the most successful description of the fundamental particles and interactions. Almost all the predictions of the SM have been verified by different experiments. Yet a number of terrestrial and extra-terrestrial observations compel us to think of the SM as a low energy theory requiring new physics at the high scale. The observation of neutrino oscillation, existence of dark matter (DM) and dark energy, baryon asymmetry of the universe are a few among the many anomalies that have been observed so far. Some of these anomalies, including the generation of light neutrino masses and the existence of a suitable DM candidate can be addressed in the Quintuplet Dark Matter model~\cite{Ko:2015uma,Agarwalla:2016rmw,Agarwalla:2018lnq} in a left-right (LR) symmetric framework that has been considered in the present work. This model is inspired by 
Minimal Dark Matter~\cite{Cirelli:2005uq,Heeck:2015qra,Garcia-Cely:2015quu,Maru:2017pwl} scenario and can also give rise to very interesting collider signatures, which we have studied in detail in this paper.

%The Standard model (SM) of particle physics is {the} most successful description of the fundamental particles and interactions. Almost all the predictions of the SM have been verified by different experiments. Yet a number of terrestrial and extra-terrestrial observations compel us to think of the SM as a low energy theory requiring new physics at the high scale. The observation of neutrino oscillation, existence of dark matter (DM) and dark energy, baryon asymmetry of the universe are a few among the many anomalies that have been observed so far. In this work, we study the collider signatures of a Minimal Dark Matter (MDM) model ~\cite{Cirelli:2005uq,Heeck:2015qra,Ko:2015uma,Garcia-Cely:2015quu,Agarwalla:2016rmw,Maru:2017pwl,Agarwalla:2018lnq} {in the framework of left-right (LR) symmetry} which can {address} some of the anomalies discussed above -- namely it can explain the small neutrino mass and has a suitable {candidate for} DM. 

Minimal Dark Matter models are minimal extensions of the SM with a bosonic or fermionic multiplet {which include a stable (over the lifetime of the universe), colored and electrically neutral particle candidate for DM}. These new multiplets are chosen to be an n-tuplet of the $SU(2)$ group with no strong interactions. The stability of the n-tuplet is either accidental or is ensured by some discrete symmetry. A quintuplet fermionic multiplet is unique in the sense that it is {the} smallest multiplet which does not require any discrete symmetries to ascertain the stability of its neutral component. {For example,} the {SM fermions and scalars being doublets or singlets under $SU(2)$, the neutral component of a $SU(2)$ quintuplet fermion} can only decay into SM particles through operators with mass dimension 6 or higher. Thus its decay width {is} suppressed at least by a factor of $1/\Lambda^2$, {where $\Lambda$ is the scale of new physics}. {In this scenario, one can ensure the stability of a} TeV scale DM {for} $\Lambda \gtrsim 10^{14}$ GeV. An $SU(2)_L$ quintuplet on the other hand has severe {constraints from observed DM relic density and limits on DM-nucleon scattering cross-section from direct DM detection experiments.} Also, due to small radiative corrections from the SM gauge bosons ($W^\pm,~Z$ and photon), the particles in an $SU(2)_L$ quintuplet are nearly degenerate (only a few hundred MeV splitting). Therefore, the decay of a component of the quintuplet to another component gives rise to soft leptons/jets making it difficult to detect at the collider experiments. However, one can exploit these small mass splittings to search for the $SU(2)_L$ quintuplet fermion particles using disappearing track signature~\cite{Ostdiek:2015aga}. $SU(2)_L$ quintuplet fermions with additional quadruplet scalars~\cite{Kumericki:2012bh,Yu:2015pwa} can also give rise to interesting signatures at the collider experiments. But, in this case, the neutral component of the quintuplet fermion is stable only over a very finely tuned region of the parameter space.

An alternate solution is to instead introduce a left-right symmetric (LRS) framework~\cite{Mohapatra:1974hk,Senjanovic:1975rk} with a quintuplet being charged under $SU(2)_R$. LRS models with the gauge group extended to $SU(3)_C\times SU(2)_L\times SU(2)_R\times U(1)_{B-L}$ are one of the 
most well motivated extensions of the SM for a number of reasons. Firstly, they can explain the origin of parity violation observed in the SM as a symmetry broken spontaneously at some high scale. The conservation of parity (P) symmetry at high scale forbids P-violating terms in the QCD Lagrangian \cite{Beg:1978mt,Mohapatra:1978fy,Babu:1989rb,Barr:1991qx,Mohapatra:1995xd,Kuchimanchi:1995rp,Mohapatra:1997su,Babu:2001se,Kuchimanchi:2010xs} and hence, can naturally solve the strong-CP problem without {introducing} a global Peccei-Quinn symmetry \cite{Peccei:1977hh}. {Moreover,} the presence of a right-handed neutrino in the particle spectrum is essentially governed by the gauge structure {and} hence, naturally explaining the origin of light neutrino masses. 

{In this work, we consider $SU(3)_C\times SU(2)_L\times SU(2)_R\times U(1)_{B-L}$ gauge symmetry and introduce a vector-like fermion multiplet which is a quintuplet under $SU(2)_R$ and singlet under $SU(2)_L$. We also introduce a scalar multiplet which is doublet under $SU(2)_R$ to break $SU(2)_R\times U(1)_{B-L} \to U(1)_Y$ and a scalar bidoublet to break the electroweak symmetry, $SU(2)_L \times U(1)_Y \to U(1)_{EW}$. In this scenario, the hypercharge quantum number is a derived quantity and allows for many different combinations of charge assignment for the quintuplet including the possibilities of having a neutral component which could be good candidate for dark matter. The dark matter phenomenology of this particular scenario was studied in details in Ref.~\cite{Agarwalla:2018lnq,Ko:2015uma}. It was shown in Ref.~\cite{Agarwalla:2018lnq} that in the presence of a singlet scalar, an $SU(2)_R$ quintuplet fermion with neutral state mass even as low as 100 GeV can explain the observed relic density (RD) data and is also consistent with DM direct detection experiments.}

{The tree-level masses of the quintuplet particles are still degenerate. However, the mass degeneracy among the quintuplet fermions of different charges are now lifted at the right-handed symmetry breaking scale (heavy right-handed gauge bosons are running in the loops) resulting in much larger mass splitting among them. In particular, the mass splitting maximizes for $(B - L) = 4$ (see Ref.~\cite{Agarwalla:2018lnq}) resulting into relatively harder leptons and jets at the collider experiments. In this work, we study the production and subsequent decay of the high charge-multiplicity components of the quintuplet for $(B - L) = 4$ at the LHC experiment. We focus our study in the same-sign lepton (SSL) channels at the 13 TeV LHC. The main advantage of SSL channels at LHC is that these channels are less background prone. The SSL channels carry distinct features of Supersymmetry and many other models as referred in several studies~\cite{Mukhopadhyaya:2010qf,Mukhopadhyay:2011xs,Bambhaniya:2013yca,Chun:2012zu}. 
ATLAS and CMS have so far looked in SSL channels in different context,
see for example Ref.~\cite{Sirunyan:2017uyt,Aad:2011vj}. We have seen
that when the neutral component's mass is small, the leptons are soft due 
to the small mass difference among the components of the quintuplets. 
Soft leptons also appear in different supersymmetric 
searches~\cite{Aad:2015mia,Aaboud:2017bac,Aaboud:2018ujj}.
Also, soft leptons in 2-opposite sign same flavor channel has been studied in 
Ref.~\cite{Sirunyan:2018iwl}, where the leading and subleading electrons 
and muons are required to satisfy $p_{T} \geq 5$ GeV.

Rest of this paper is organized as follows. In Section 2, we briefly 
introduce the model. Production and subsequent decays of the charged 
components of the quintuplet fermions are discussed in Section 3. 
In Section 4, we discuss the two, three, and four same-sign multilepton 
signatures (2SSL, 3SSL, and 4SSL) and the corresponding SM 
backgrounds at the LHC. We also present the exclusion limit 
and discovery reach at the 13 TeV LHC for SSL channels in detail. 
Finally, we conclude in Section 5 with discussions.}

%============================
\section{Quintuplet Dark Matter Model}
\label{model}
%============================

The gauge group in left-right symmetric models is extended 
to $SU(3)_C \times SU(2)_L \times SU(2)_R \times U(1)_{B-L}$. The matter fermion sector is given as:
\begin{eqnarray}
&Q_L& \left(3,2, 1, \frac13 \right ) = \!\left (\begin{array}{c}
u\\ d \end{array} \right )_L,~~~~
Q_R\left ( 3,1, 2, \frac13
\right )\!=\!\left (\begin{array}{c}
u\\d \end{array} \right )_R,\nonumber \\ [4pt]
&l_L&\left ( 1,2, 1, -1 \right )=\left (\begin{array}{c}
\nu\\ e\end{array}\right )_L,~~~~
l_R\left ( 1,1, 2, -1 \right )=\left (\begin{array}{c}
\nu \\ e \end{array}\right )_R,
\end{eqnarray}
where, the numbers in the bracket corresponds to $SU(3)_C,~ SU(2)_L,~SU(2)_R~{\text{and}}~U(1)_{B-L}$ quantum numbers respectively. Here we see that all the quarks and leptons are a part of either left-handed or right-handed doublets. For any particle in this model, the electric charge $Q$ is given as: $Q=T^3_L+T^3_R+Q_{(B-L)}/2$, where $T^3_{L/R}$ represents the third component of the isospin for $SU(2)_{L/R}$. An additional singlet fermion $N$(1,1,1,0) is also introduced to generate the light neutrino mass through inverse seesaw mechanism \cite{Minkowski:1977sc,Yanagida:1979as,Sawada:1979dis,Levy:1980ws,VanNieuwenhuizen:1979hm,Mohapatra:1979ia}.

To break the right-handed symmetry, electroweak (EW) symmetry and to generate the quark and lepton masses and mixing, a minimal scalar Higgs sector is required and given by,
%=========
\begin{eqnarray}
H_R(1,1,2,1)&=&\left (\begin{array}{c}
H_R^+ \\H_R^0 \end{array} \right ),~~~~
\Phi(1,2,2,0)={\left (\begin{array}{cc}
\phi^{0}_1 & \phi^{+}_{2} \\ \phi^{-}_{1} & \phi^{0}_{2} \end{array} \right)}.~~~
\label{scalar}
\end{eqnarray}
%==========
The non-zero vacuum expectation values (VEV) of the scalar fields are $\left< H^0_R \right> = v_R$, $\left< \phi_{1}^0 \right> = v_1$ and $\left< \phi^0_{2} \right> = v_{2}$. The VEV of the doublet $v_R$ is responsible for breaking the right-handed symmetry while bidoublet VEVs $v_1$ and $v_2$ break the EW symmetry.
%%%%%%%%%%%%%%%%%%%%%%%%%%%%%%% 
 
{One can thus obtain the charged gauge boson mass-squared matrix ($M_W^2$) in the basis $(W_R^\pm, W_L^\pm)$ and the neutral gauge boson mass-squared matrix ($M_Z^2$) in the basis $(W^3_R, W^3_L, V)$ as:
%==============
\begin{equation}
M_W^2 = \frac{1}{2}\begin{bmatrix}
g_R^2\left( v_R^2+v^2 \right)&g_L g_R v_1 v_2 \\ g_L g_R v_1 v_2&g_L^2 v^2
\end{bmatrix},~~~~
M_Z^2 = \frac{1}{2}\begin{bmatrix}
g_R^2\left(v_R^2+v^2\right)&g_L g_R v^2&-g_R g_{B - L} v_R^2 \\ g_L g_R v^2&g_L^2 v^2&0 \\-g_R g_{B - L} v_R^2&0&g_{B - L}^2 v_R^2
\end{bmatrix},
\end{equation}}
%===============
%%%%%%%%%%%%%%%%%%%%%%%%%%%%%%%%%%
where, $v^2=v_1^2+v_2^2$ is the EW VEV $\sim$ 174 GeV while $g_L$, $g_R$ and $g_{B-L}$ are the $SU(2)_L$, $SU(2)_R$ and $U(1)_{B-L}$ gauge couplings respectively. Our model corresponds to a LRS scenario where the parity and right-handed symmetry breaking scales are decoupled \cite{Chang:1983fu}. This is evident from the fact that our scalar spectrum does not contain a left-handed doublet, and hence the left and right-handed gauge couplings may not be the same.  
Thus, neglecting the left-right mixing, the new right-handed heavy gauge bosons masses are :
\begin{eqnarray}
M^2_{W_R}&=&\frac{1}{2}g_R^2\left( v_R^2+v^2 \right),~~~~M^2_{Z_R} = \frac{1}{2}\left(g_R^2+g^2_{B - L}\right) \left[ v_R^2+\frac{g_R^2 v^2}{\left(g_R^2+g^2_{B - L}\right)} \right].
\end{eqnarray}
The left-handed $W$ and $Z$ boson masses are given by their usual expression in the SM 
with the identification of the effective hypercharge gauge coupling as,
\begin{eqnarray}
g_Y = \frac {g_R~g_{B - L}}{\sqrt{(g_R^2+g_{B - L}^2)}}. 
\end{eqnarray}
The right-handed VEV needs to be quite high so as to get a large enough $Z_R$ mass in order to avoid direct detection constraints for dark matter.
As can be seen from Fig. 5 of Ref.~\cite{Agarwalla:2018lnq}, the mass of the $Z_R$ boson should be at least 7 TeV to circumvent the direct detection constraints for a dark matter mass of 150 GeV. Since we have considered a parameter space with the lowest value of quintuplet dark matter mass of 150 GeV for our collider analysis, this limit needs to be satisfied. For this reason, we choose $v_R$ = 13 TeV, which gives $M_{W_R}$ = 6.0 TeV and $M_{Z_R}$ = 7.14 TeV.
%%%%%%%%%%%%%%%%%%%%%%%%%%%%%%%%%%%%%%%%

Motivated by Minimal Dark Matter models, we consider an $SU(2)_R$ vector-like fermion quintuplet, which can accommodate the dark matter. 
We represent the quintuplet as,
\begin{eqnarray}
%\Sigma(1,1,5,X) = ( \chi^{2+X/2},\chi^{1+X/2},\chi^{X/2},\chi^{-1+X/2},\chi^{-2+X/2})^T,
\chi(1,1,5,X) = ( \chi^{2+X/2},\chi^{1+X/2},\chi^{X/2},\chi^{-1+X/2},\chi^{-2+X/2})^T,
\end{eqnarray}
where $X=0,2,4$ are its possible $(B - L)$ quantum numbers. The neutral component of {$\chi$} 
can be a viable dark matter while its charged components can be produced at the colliders. 
The couplings of the gauge bosons ($Z$, $Z_R$, $W_R$ and photon) with the quintuplet fields are,
%===============
\begin{eqnarray}
\mathcal{L}_{gauge} \supset &-& \frac{g_Y^2}{\left(g_R^2+g_Y^2\right)} Q_{\chi^i} \overline{\chi}^i Z^\mu~\gamma_\mu~\chi^i + e~Q_{\chi^i} \overline{\chi}^i A^\mu \gamma_\mu~\chi^i \notag \\
&+& \sqrt{g_R^2 - g_Y^2} \left[Q_{\chi^i} - \frac{g_R^2 Q_{B - L}}{2 \left(g_R^2 - g_Y^2\right)} \right]\overline{\chi}^i Z_R^\mu \gamma_\mu \chi^i + (\frac{g_R}{\sqrt{2}} r_Q \ov\chi^{i+1}~W^\mu_R~\gamma_\mu~\chi^i + h.c.).~~~~~
\label{eq:gaudm}
\end{eqnarray}
%===============
Here 
\begin{equation}
r_Q = \sqrt{(3+Q_{\chi_i}-Q_{B - L}/2)(2-Q_{\chi_i}+Q_{B - L}/2)}
\end{equation}
and $\chi^i$ represents the {component of} the quintuplet {$\chi$} with electric charge $Q_{\chi^i}{=i}$.

The fermion masses are generated as the bidoublet fields get non-zero VEV and the corresponding Yukawa Lagrangian is given as:
%==============
\begin{eqnarray}
\mathcal{L}_Y&=& \left(Y_q\overline{Q}_{L}\Phi Q_{R}+\widetilde{Y}_q\overline{Q}_{L}\widetilde{\Phi}Q_{R}+Y_l\overline{l}_{L}\Phi l_{R}+\widetilde{Y}_l\overline{l}_{L}\widetilde{\Phi}l_{R} + f_{R}\overline l_{R} \widetilde H_R N + H.C. \right) \notag \\ 
&+& \frac{{\mu_N}}{2} N N + M_\chi \overline{\chi} \chi~~,
\label{eq:yukawa}
\end{eqnarray}
%===============
where $Y$ and $f$ are the Yukawa couplings and $\widetilde{\Phi}=\tau_2\Phi^\ast\tau_2,~\widetilde H_{R} = i \tau_2 H^\ast_{R}$.
Hence, the quark and charged lepton masses are:
%%%%%%%%%%%%%%%%%%%%%%%%%%%%%%%%%%%%%%%%%%%%%%%%%%%%%%%%%%===========
{\vspace*{-0.2cm}
\begin{eqnarray}
M_{u} &=& Y_q v_1+\wt{Y}_q v_2,~~M_d=Y_q v_2+\wt{Y}_q v_1,~~M_l = Y_l v_2+\wt{Y}_l v_1.~~~~
\label{eq:fermass}
\end{eqnarray}
%===========
We choose a large $\tan{\beta}~(=v_1/v_2)$ limit so that $Y_{33}^q \sim 1$ for the top quark mass while $\wt{Y}_{33}^q < 10^{-2}$. A smaller value of $\tan \beta$ would require a larger value of $Y_{33}$ in general, leading to the top Yukawa becoming non-perturbative at relatively lower mass scales. The neutrinos get masses through the inverse seesaw mechanism with the $3\times3$ mass matrix in the basis $(\nu_L,\nu_R,N)$ given as:
\begin{equation}
M_{\nu}=\begin{bmatrix}
0&m_D&0\\m_D^T&0&f_R v_R\\0&f_R^T v_R&\mu_N
\end{bmatrix},
\end{equation} 
where $m_D = Y_l v_1+\wt{Y}_lv_2$ is the neutrino Dirac mass term. Assuming $f_{R} v_{R} >> m_D, \mu_N$, the approximate expressions for the neutrino mass eigenvalues (for one generation) are,
\begin{equation}
m_{\nu_1} \sim \frac{(f_R^{-1} M_D^T)^T \mu_N (f_R^{-1} M_D^T)}{v_R^2}, ~~~~m_{\nu_{2,3}} \sim f_R v_R.
\end{equation}
So in this framework, light neutrino masses can be easily generated by appropriate choice of parameters.}

\begin{figure}[!ht]
\centering
\includegraphics[width=7.7cm]{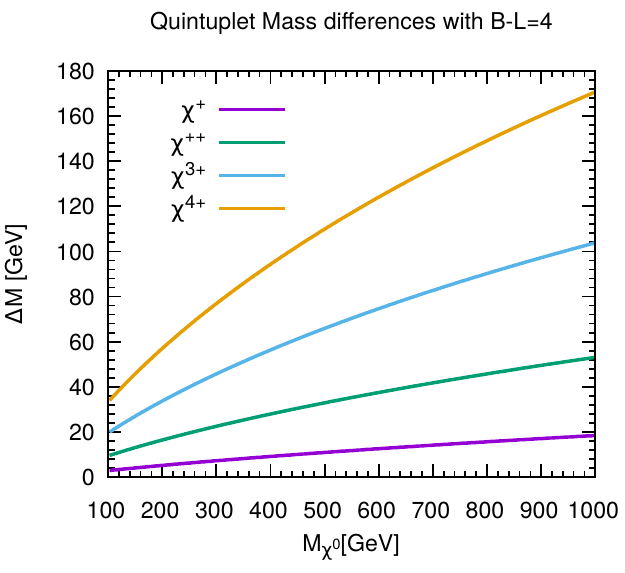}
\mycaption{Mass difference between various charged states as a function of the neutral state's mass 
for\\$(B - L) = 4$ case, assuming $M_{W_R}$= 6 TeV and $M_{Z_R}$= 7.14 TeV (see Eq.~\ref{eq:mass}).}
\label{fig:msplit} 
\end{figure}

The quintuplet fermion in this model can have a few different values of $(B - L)$ quantum numbers (0, 2, and 4) but for our study we will only consider the case with $(B - L)$ = 4. This scenario will have highest charge multiplicity particles, hence gives rise to interesting collider signatures. Component fields of the $SU(2)_R$ quintuplet for $(B - L)$ = 4 can be expressed as,
\begin{equation}
\chi(1,1,5,4) = \left(\chi^{++++},\chi^{+++},\chi^{++},\chi^+,\chi^0\right)^T.
\end{equation}
For brevity, we will denote $\chi^{\pm\pm\pm\pm}$ as $\chi^{4\pm}$, $\chi^{\pm\pm\pm}$ as $\chi^{3\pm}$, and $\chi^{\pm\pm}$ as $\chi^{2\pm}$ from here onwards. It is evident from Eq.~\ref{eq:yukawa}, that all {components} ($\chi^i$) {of} the quintuplet are degenerate in mass at the tree level. Their masses are all equal to $M_\chi$ as given in Eq.~\ref{eq:yukawa}. The mass splitting between the various charged states are generated by the radiative corrections given as:
%============
\vspace*{-0.2cm}
\begin{eqnarray}
M_{\chi^i}-M_{\chi^0}&=&\frac{g_R^2}{(4\pi)^2}M_{\chi^0} {\bigg{[}} Q_{\chi^i}\left(Q_{\chi^i}-Q_{B - L}\right)f(r_{W_R}) - Q_{\chi^i} \left(\frac{\sqrt{g_R^2-g_Y^2}}{g_R^2}Q_{\chi^i}-Q_{B - L} \right)f(r_{Z_R})\nonumber \\
&-& \left. \frac{g_Y^2}{g_R^2}Q_{\chi^i}^2\left\{s_W^2 f(r_{Z}) + c_W^2 f(r_{\gamma}) \right\} \right],
\label{eq:mass}
\end{eqnarray}
%=============
where $r_X$ = $m_X/M_{\chi^0}$, $f(r) \equiv 2 \int_0^1 dx (1+x) \text{log} \left[ x^2 + (1-x)r^2 \right]$ and $Q_{\chi^i}$ is the electric charge of $\chi^i$. To calculate the mass splittings, we have chosen the value of $g_R =0.653$ at the scale of $Z_R$ boson mass. This is merely a choice in our parameter region. In the limit of small left-right mixing, the value of $g_L=0.653$ at the electroweak scale is determined by the measured values of $\alpha_{EM}$ and weak mixing angle $\theta_W$. The mass splitting between the various charged states 
of the quintuplet as a function of the neutral quintuplet 
mass has been shown in Fig.~\ref{fig:msplit} for the case 
with $(B - L)$ = 4. The mass difference between the components of the 
quintuplets increase with their mass. In order to satisfy the correct RD, one 
needs to introduce a singlet scalar as shown in 
Ref.~\cite{Agarwalla:2018lnq}. The same paper also 
has in-depth discussion of the model, dark matter 
phenomenology and collider phenomenology 
of the singlet scalar. In this work, we are interested 
in the collider signatures of the quintuplet fermions 
at the LHC which will be discussed in the following sections.
%===============================
%================================
%===================================================
\section{Production and Decay of Multi-charged Quintuplet Fermions}
%===================================================

\begin{figure}[!ht]
\centering
\includegraphics[width=7.0cm]{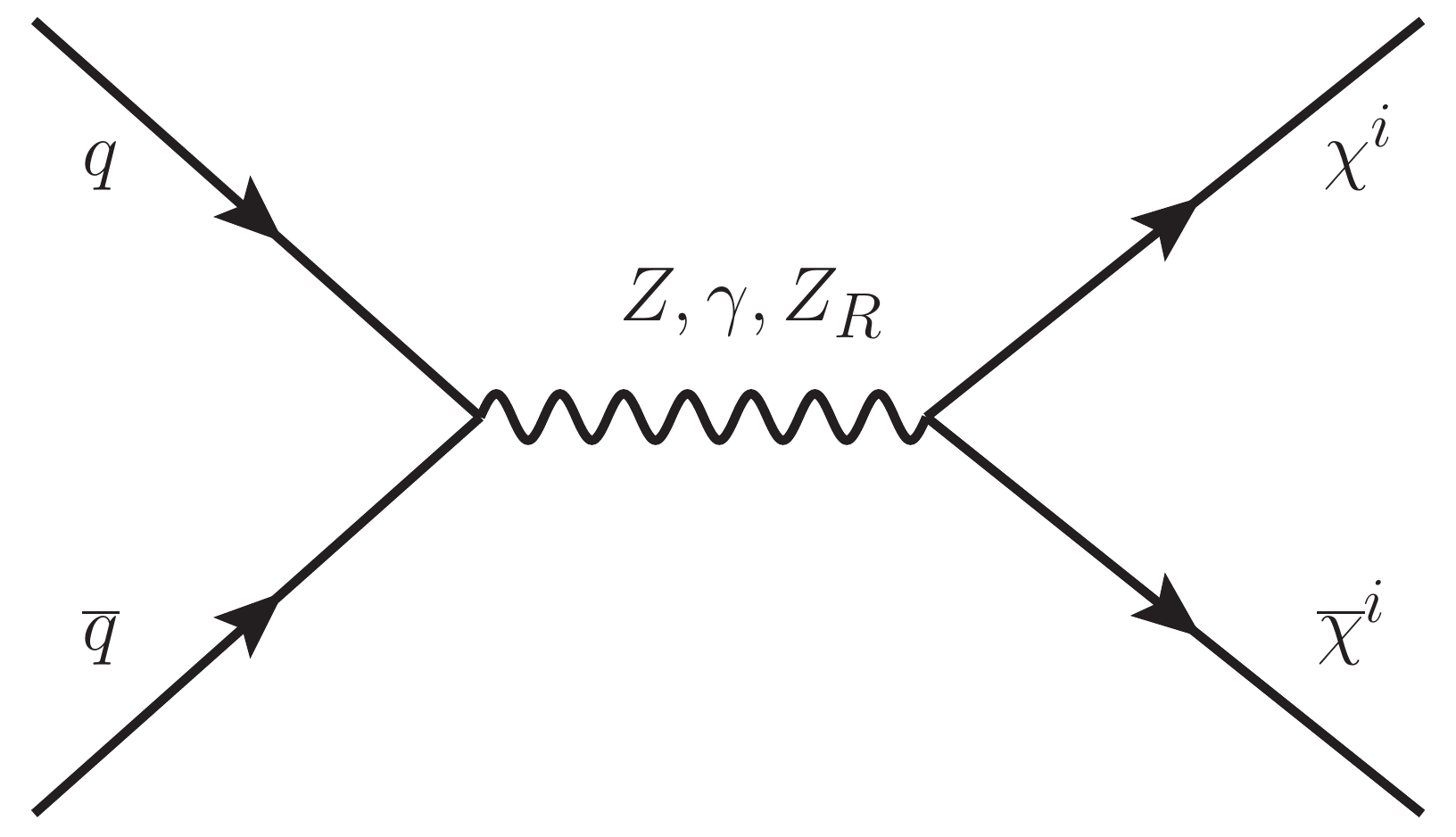}
\includegraphics[width=7.0cm]{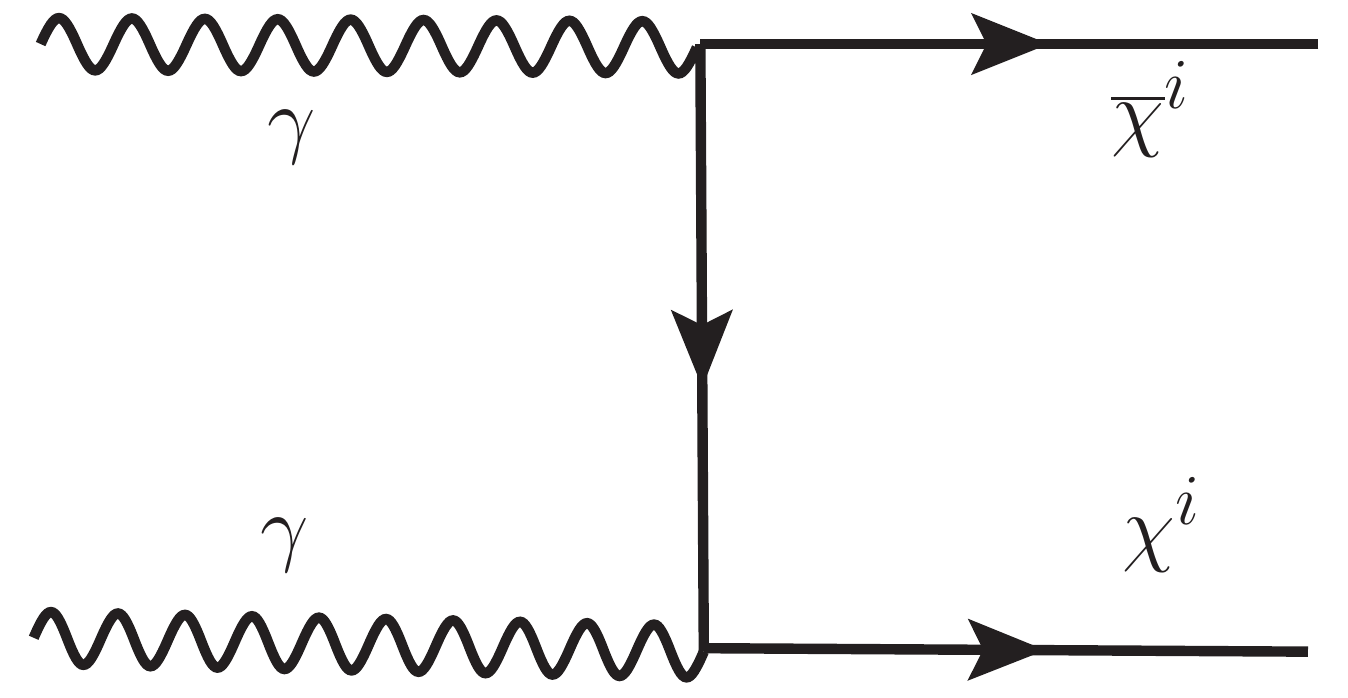}
\mycaption{The left panel shows the Feynman diagram 
for the s-channel Drell-Yan pair-production of the 
quintuplet fermion. The right panel portrays the
Photo production of $\chi^{i} \overline\chi^{i}$ 
through t-channel photon-photon fusion process.}
\label{fig:feyn}
\end{figure}

The quintuplet, being charged under $SU(2)_R$ and $U(1)_{B-L}$, 
has gauge interaction with photon, SM $Z$ boson, $Z_R$ and $W^{\pm}_R$ 
(see Eq.~\ref{eq:gaudm})\footnote{The couplings of the quintuplet fermions with 
the SM $W^\pm$ is suppressed by small $W^\pm_L$--$W^{\pm}_R$ mixing.}. 
Therefore, the quintuplet fermions can be pair-produced at the LHC 
via quark-antiquark initiated Drell-Yan (DY) process with a photon/$Z$/$Z_R$ 
in the $s$-channel as shown in the left panel of Fig.~\ref{fig:feyn}. The electrically 
charged components of the quintuplet ($\chi$) can also be produced from photon-photon fusion 
in the initial state ($\gamma \gamma \to \chi^i \overline{\chi}^i$), 
where $i$ = 4, 3, 2, and 1. The right panel of Fig.~\ref{fig:feyn} shows a representative 
diagram of the photon-photon fusion process for the pair-production of 
$\chi^i \overline{\chi}^i$. Photo production of $\chi^{i\pm}\chi^{i\mp}$ pairs 
takes place via a $t$ or a $u$ channel exchange of a $\chi^{i\pm}$ and 
hence, is not suppressed by the parton center of mass energy. 
Moreover, the coupling of photon with a pair of $\chi^{i\pm}$ 
being proportional to its charge (i), the matrix element 
squared of photo productions are enhanced by a factor of $i^4$. 
However, the photo-production of charged fermions at the LHC is suppressed by the small parton density of photon inside a proton.
To denote the charge multiplicity of the quintuplet 
in the following, we adopt the notation $\chi^{n\pm}$, where $n$ runs from 0 to 4. 

%=============================
%===============================

In fact, the parton density of the photon is so small that most of the older versions of PDF's do not include photon as a parton. However, if we want to include QED correction to the PDF, inclusion of the photon as a parton with an associated parton distribution function is necessary. In the era of precision physics at the LHC when PDF's are determined upto NNLO in QCD, NLO QED corrections are important (since $\alpha_{s}^2$ is of the same order of magnitude as $\alpha$) for the consistency of calculations. Moreover, as discussed previously, photon-initiated processes could become significant at high energies for some processes. In view of these facts, NNPDF \cite{Ball:2014uwa,Ball:2013hta}, MRST \cite{Martin:2004dh} and CTEQ \cite{Schmidt:2015zda} have already included photon PDF into their PDF sets.

\begin{figure}[!ht]
\centering
\includegraphics[width=7.7cm]{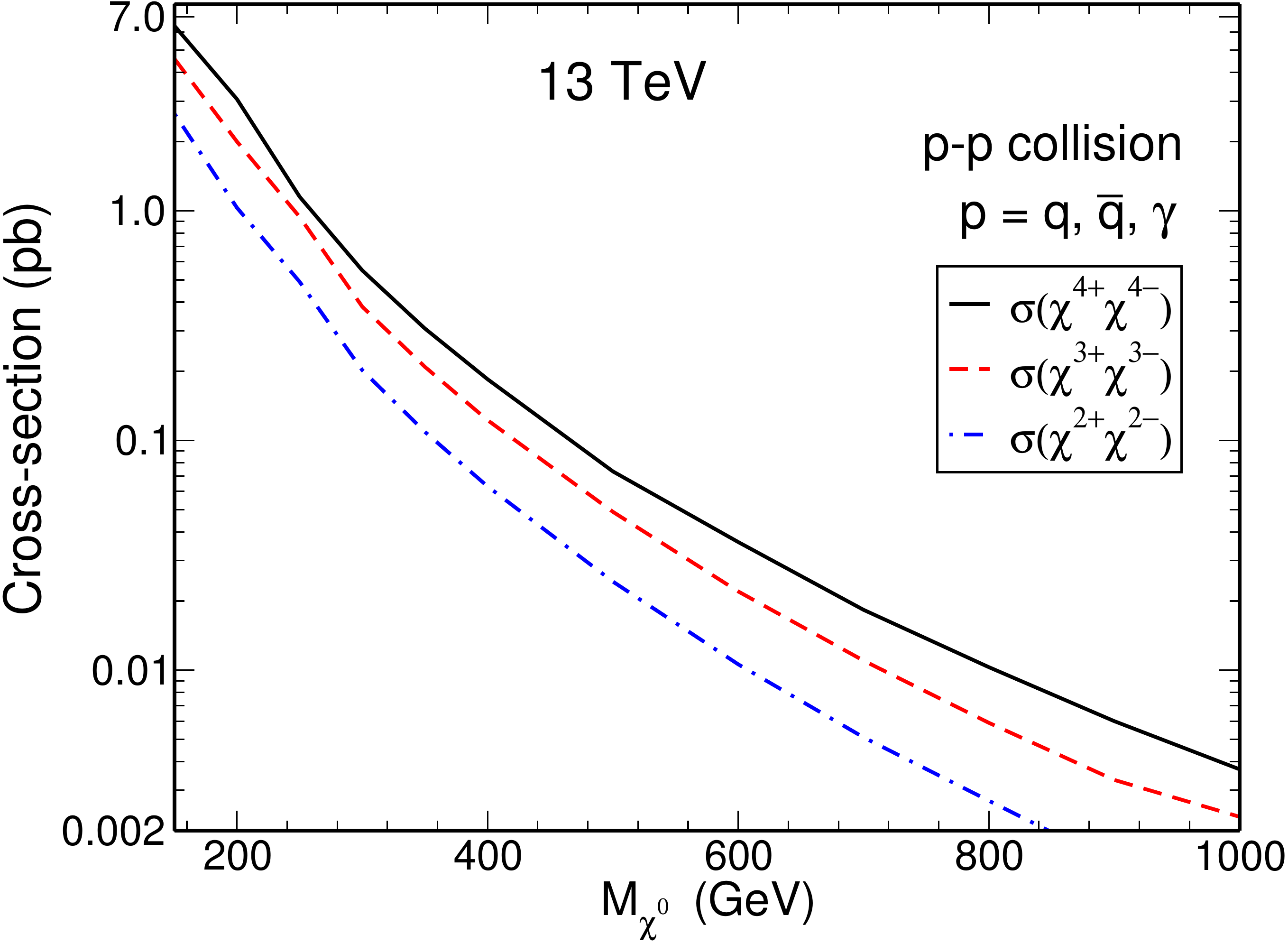}
\includegraphics[width=7.7cm]{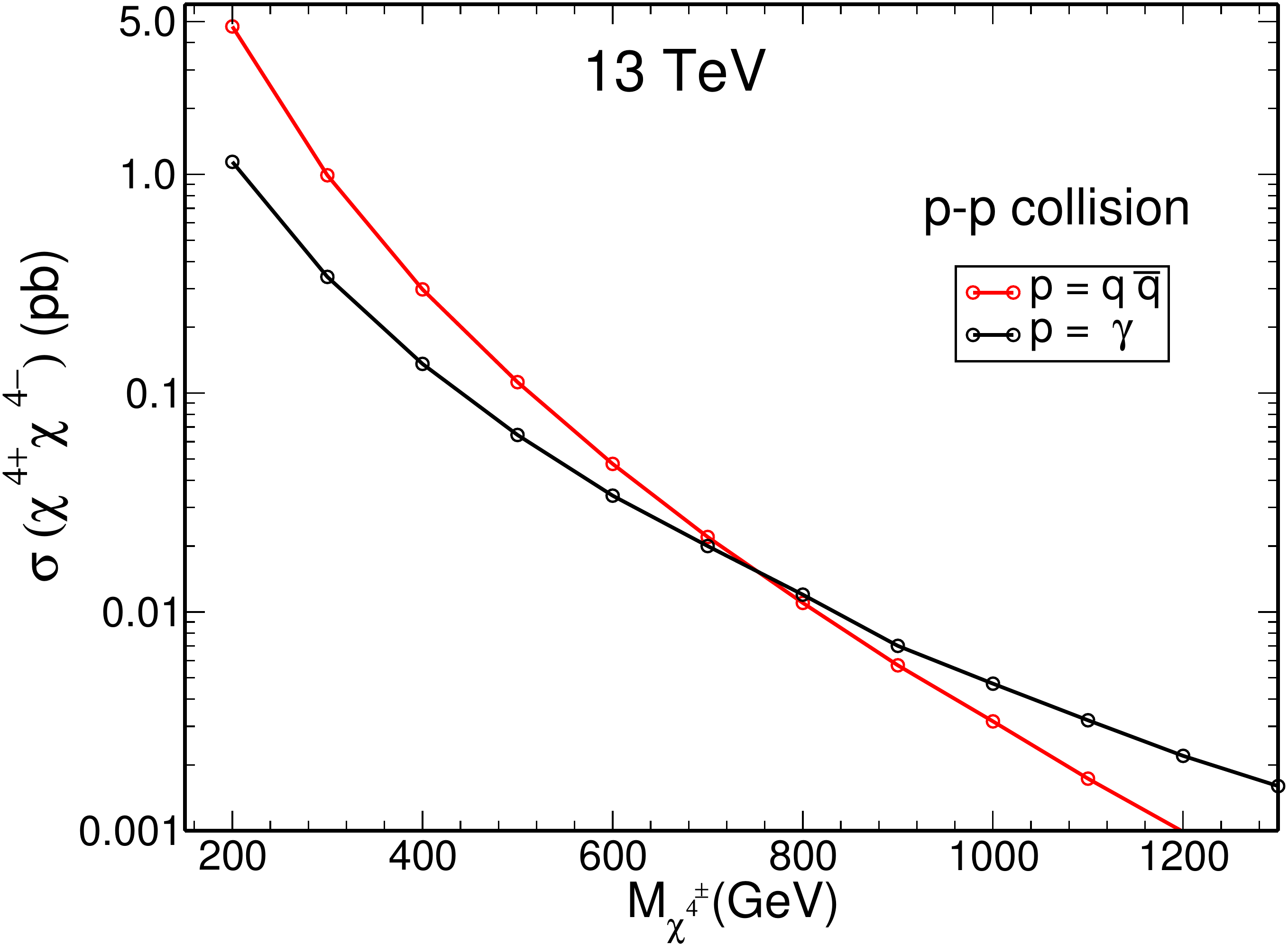}
\mycaption{Left panel shows the total cross-section of different production modes 
as a function of the neutral component's mass ($M_{\chi^0}$) at 13 TeV when both 
quarks and photons are considered at the initial states. The cross-section of other 
production modes are negligible. In the right panel, we give the 
cross-sections to produce a pair of quadruply charged 
particles as a function of its mass ($M_{\chi^{4\pm}}$) at 13 TeV. We show it separately for two different 
initial states (quark-antiquark and photons).}
\label{cross} 
\end{figure}

In order to compute the cross-sections and generate events at the LHC, we incorporate the model Lagrangian of Eq.~\ref{eq:gaudm} in FeynRules (v2.3.13) \cite{Alloul:2013bka,Christensen:2008py}. We implement the couplings for each component of the quintuplet ($i$ =0, 1, 2, 3, 4) with the gauge bosons, derived from Eq.~\ref{eq:gaudm}. Using FeynRules we generate the model file for MadGraph5\_aMC@NLO (v2.2.1)~\cite{Alwall:2014hca}. For the cross-sections, we use {the} NNPDF23LO1 parton distributions~\cite{Ball:2012cx} with the factorization and renormalization scales kept fixed at the central $m_T^2$ scale after $k_T$-clustering of the event. The quark-antiquark initiated DY production cross-section for the $\chi^{4+}\chi^{4-}$ pairs are presented in Fig.~\ref{cross} (right panel) at the LHC with 13 TeV center of mass energy. Being $s$-channel, DY pair-production cross-sections are significantly suppressed for larger $\chi^{4\pm}$ masses. In Fig.~\ref{cross} (right panel), we also present the photo-production cross-section of $\chi^{4\pm}\chi^{4\mp}$ pairs as a function of $\chi^{4\pm}$ mass ($M_{\chi^{4\pm}}$) at the 13 TeV LHC. It shows that photon-photon fusion contributes significantly for $M_{\chi^{4\pm}}<800$ GeV and for $M_{\chi^{4\pm}}>800$, photo-production dominates over the DY contribution. In the left panel of Fig.~\ref{cross} we present the total (DY+photo-production) pair-production cross-sections for $\chi^{4\pm}\chi^{4\mp}$, $\chi^{3\pm}\chi^{3\mp}$, and $\chi^{2\pm}\chi^{2\mp}$ as a function of $M_{\chi^{0}}$ at the LHC with $\sqrt s=13$ TeV. The total pair-production cross-sections varies between a few pb to a few fb as we vary the $\chi^{0}$ mass between 200 GeV to 1 TeV. Quintuplet fermions can also be produced in association with another quintuplet fermion from quark-antiquark initial state via a $W^{\pm}_R$ exchange in the $s$-channel. However, associated production cross-sections are suppressed by the mass of the $W^{\pm}_R$ in the $s$-channel. In Table~\ref{table1}, we present the numerical values of pair and associated production cross-sections for a particular benchmark point.

%====================================================
%=====================================================

%=====================================================
\begin{table}[!ht]
\centering
\begin{tabular}{|c|c|c|c|}
\hline\hline
Process & Cross-section (pb) & Process & Cross-section (pb) \\
\hline
$p p \rightarrow \chi^{4+}\chi^{4-}$ & 0.1843 & $p p \rightarrow \chi^{4+}\chi^{3-}$ & 0.00058 \\
$p p \rightarrow \chi^{3+}\chi^{3-}$ & 0.1225 & $p p \rightarrow \chi^{3+}\chi^{2-}$ & 0.00088 \\
$p p \rightarrow \chi^{2+}\chi^{2-}$ & 0.0630 & $p p \rightarrow \chi^{2+}\chi^{1-}$ & 0.00078 \\
$p p \rightarrow \chi^{1+}\chi^{1-}$ & 0.0174 & $p p \rightarrow \chi^{1+}\chi^{0}$ & 0.00052 \\
\hline\hline
\end{tabular}
\mycaption{Cross-sections at the 13 TeV LHC in different production channels 
when $M_{\chi^{n\pm}}$ = [400,409,428,456,494] GeV ($n$ = 0 to 4), and initial state 
partons and photons both are considered, $p$ = $\gamma$, $q$, 
$\overline{q}$. The contribution from $p p \rightarrow \chi^{0}\chi^{0}$ 
is negligible.}
\label{table1}
\end{table}
%========================================================

\begin{figure}[!htb]
\centering
\includegraphics[width=0.5\textwidth]{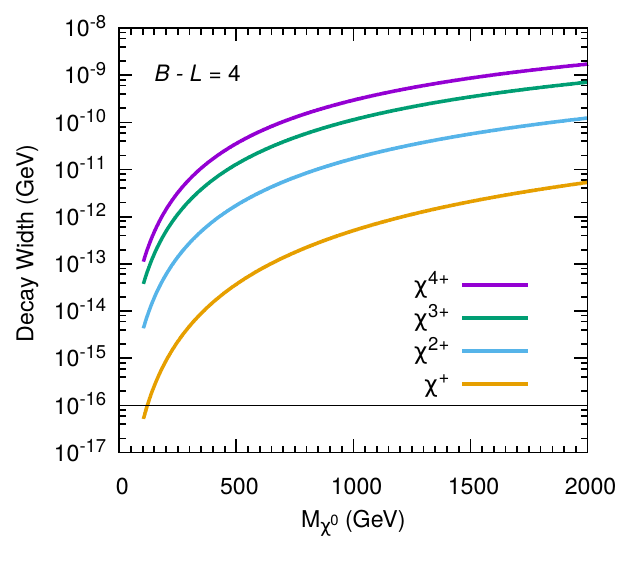}
\mycaption{Total decay width of the charged states a function of the neutral state's mass 
for $(B - L) = 4$ case, assuming $M_{W_R}$= 6 TeV and $M_{Z_R}$= 7.14 TeV. If the decay width 
is less than $10^{-16}$ GeV, the particle decays outside the detector.}
\label{width}
\end{figure}

After being produced at the LHC, the quintuplet fermions undergo 
a tree-level 3-body decay into a lighter component and a pair of SM 
quarks or leptons. $\chi^{0}$ being stable remains invisible in the detector. 
Therefore, the pair/associated production of the quintuplet fermions 
gives rise to multiple-jets and/or leptons (including the same-sign multileptons) 
and missing transverse energy signature at the LHC. However, the 3-body 
decay of the quintuplet fermions proceed through an off-shell $W^{\pm}_R$ boson 
with a mass of few TeV. Therefore, before going into the details of signal and 
background analysis, it is important to compute the decay width of the quintuplet 
fermions to ensure that they decay inside the detector. For a particle having a decay width 
less than $10^{-16}$ GeV, it will escape the detector before it can decay. 
We can see from Fig.~\ref{width} that the decay widths of all the charged fermions 
in the $(B - L) = 4$ quintuplet are always larger than $10^{-16}$ GeV, which ensures that they will 
decay inside the detector. If the decay width of the charged fermions fall in the range 
around $10^{-13}$ to $10^{-16}$ GeV, then there may be a possibility to see the displaced 
vertex signature as well. For a major part of the parameter space considered in this paper, 
the decay widths of $\chi^{2\pm}$, $\chi^{3\pm}$, and $\chi^{4\pm}$ are always above $10^{-13}$ GeV (see Fig.~\ref{width}). 
Therefore, we do not explore the possibility of displaced vertex signature in this work. 

%=========================================
%==========================================

\begin{figure}[!htb]
\centering
\includegraphics[width=0.7\textwidth]{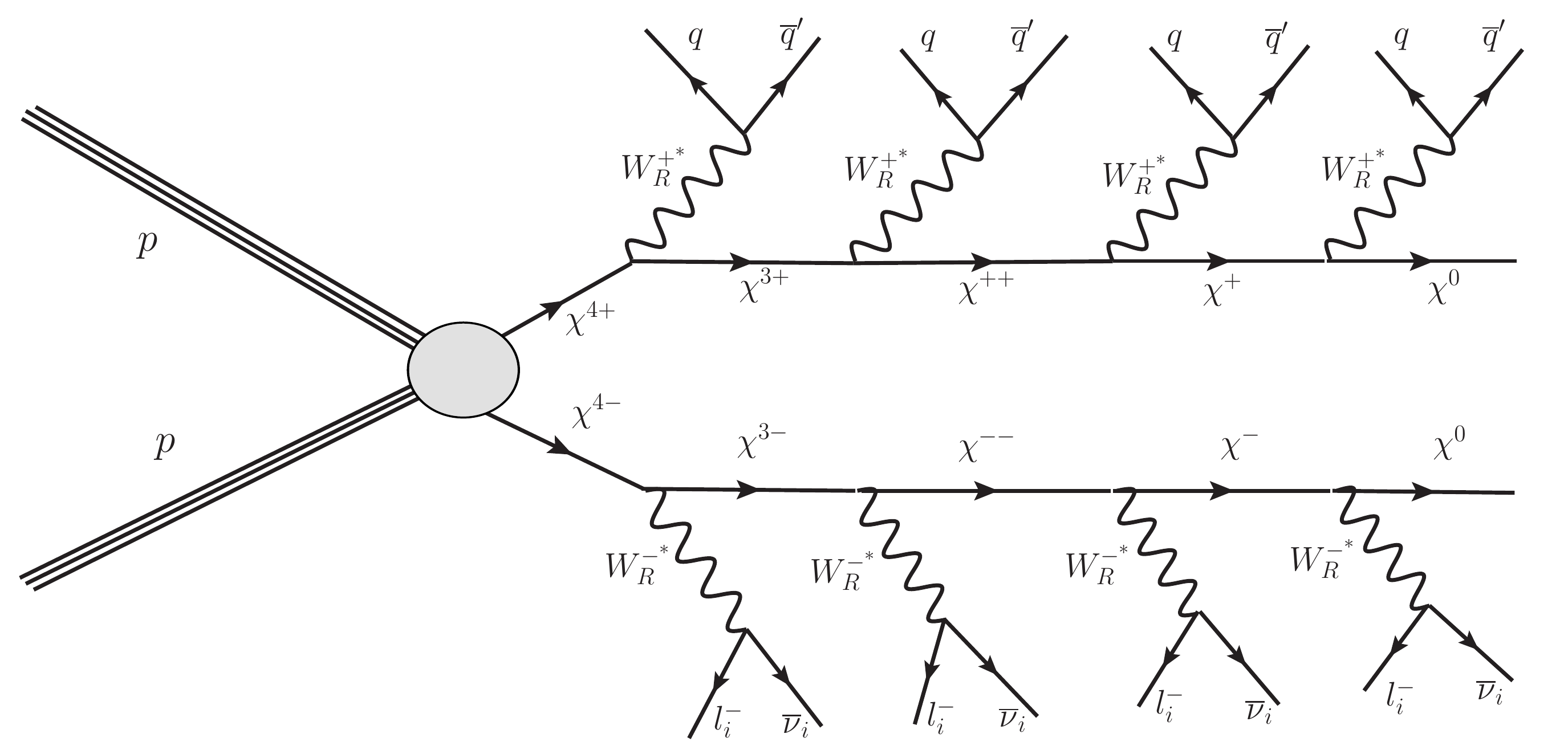}
\mycaption{Cascade decay of $\chi^{4\pm}$ into leptons ($l=e,\mu$) and jets, contributing to the 4SSL channel.}
\label{cascade}
\end{figure}

If a charged ($n$ = 1, 2, 3, 4) component of the quintuplet ($\chi^{n\pm}$) 
is produced, it will form a cascade of decay via the process discussed above. 
For example, the main contribution in the 4 same-sign lepton channel (4SSL) 
will come from the production $p p \rightarrow \chi^{4+}\chi^{4-}$, 
and the decays of $\chi^{4+}$ and $\chi^{4-}$, as shown in
Fig.~\ref{cascade}. Other components of the quintuplet can also decay in the same manner and 
depending on the decay pattern there can 
be interesting collider signatures in the same-sign lepton (SSL)
and other multilepton channels. In this study, we focus 
mainly on 2, 3, and 4 same-sign multilepton signals (2SSL, 3SSL, and 4SSL). 

%=========================================
%==========================================

%===============================
\section{Collider Signature of Quintuplets}
%===============================

As discussed in the previous section, the production and decay 
of the quintuplet fermions give rise to multiple leptons/jets in association 
with missing transverse energy signatures at the LHC. For example, 
the production of $\chi^{4\pm}\chi^{4\mp}$ pairs could result into 
0 to 8 leptons (including same-sign 2, 3, and 4 leptons) in the final state 
depending on the decay cascade of $\chi^{4\pm}$. Fully leptonic/hadronic 
decay cascade of both $\chi^{4\pm}$ results into 8/0 leptons signature. 
Whereas, leptonic decay cascade (fully or partially) for one $\chi^{4\pm}$ 
and hadronic decay cascade of the other $\chi^{4\mp}$ give rise to four same-sign 
leptons in the final state. The pair and associated production of all 
combinations of $\chi^{4\pm}, \chi^{3\pm},~{\rm and}~\chi^{2\pm}$ 
contribute in the 2SSL channel. For the 3SSL channel, $\chi^{4+}\chi^{4-}$, $\chi^{3+}\chi^{3-}$, 
$\chi^{4+}\chi^{3-}$, and $\chi^{3+}\chi^{2-}$ contribute, but for the 4SSL channel the signal stems only from 
$\chi^{4+}\chi^{4-}$ and $\chi^{4+}\chi^{3-}$. 
The cross-section ($\sigma$) $\times$ effective Branching Ratio (BR) in different SSL channels are 
listed in Table~\ref{table2} for a few selected masses of the quintuplet. 
The simulation of production and decay of the quintuplet fermion pairs 
at the LHC are discussed in the following section.

%=================================
\begin{table}[ht!]
\centering
\begin{tabular}{|c|c|c|c|}
\hline\hline
$M_{\chi^{n\pm}}$ (GeV) & $\sigma \times$BR(fb)(2SSL)& $\sigma\times$BR(fb)(3SSL)& $\sigma\times$BR(fb)(4SSL)\\
\hline
150,154,163,177,195 & 886.5 & 132.54 &24.17 \\
200,205,216,233,257 & 214.4 &71.88&11.55 \\
250,256,269,289,317 & 147.1 & 32.45&6.0 \\
300,307,322,345,377 & 75.83 &11.23&2.0 \\
350,358,375,401,436 & 41.35 &6.21&1.16 \\
400,409,428,456,494 & 24.73 &3.7&0.7 \\
\hline\hline
\end{tabular}
\mycaption{Signal cross-sections (production cross-section $\times$ effective BR) in the 2SSL, 3SSL, and 4SSL channels at the 13 TeV LHC for different masses of the quintuplet. Here {\it n} runs from 0 to 4 for a given quintuplet. This cross-sections are obtained considering both quarks and photons in the initial state. We consider BR($\chi^{Q+1} \rightarrow l \nu \chi^{Q}$) = 0.22~($l=e,\mu$), and 
BR($\chi^{Q+1} \rightarrow q \overline{q} \chi^{Q}$) = 0.67.} 
\label{table2}
\end{table}
%===================================

%=====================
\subsection{Event Generation and Simulation}
%======================

The signal events for the production of the quintuplets are simulated using 
MadGraph \cite{Alwall:2011uj, Alwall:2014hca} and showered with 
PYTHIA \cite{Sjostrand:2006za}. After that, the events are
passed through DELPHES~3 \cite{deFavereau:2013fsa} for detector 
simulation. In DELPHES, we choose the isolation cut for leptons to 
be $\Delta R_{max} = 0.5$ while reconstructing the events. 
This requirement ensures no hadronic activity inside this isolation cone. 
The isolation cut reduces the SM background of the leptons coming from the decays of B-mesons. 
The probability of a jet to be misidentified as a lepton is taken as a modulo inside DELPHES~\cite{Khachatryan:2014sta}. In our model, leptons get produced from the 
decay of a heavier component of the quintuplet to a lighter component. As discussed earlier, small mass splitting between different components of the quintuplet results in soft leptons. For such soft leptons the charge misidentification probability 
is small and hence neglected. 

The leading SM backgrounds for 2SSL are $t\overline{t}$, 
$ZZ$, $WZ$, $WW$, $t\overline{t} W$, $t\overline{t} Z$, $t\overline{t} h$, 
and $hZ$. Semileptonic decay of $t\overline{t}$ contributes to the 2SSL 
when the b-quark generated from decay of top quark 
further decays leptonically. On the other hand, di-boson production 
contributes to the 2SSL channel if one or more leptons from the leptonic decay 
of the SM bosons fall outside the detector coverage (leptons are too soft or they fall 
in the high rapidity region). Here, 
$b\overline{b}W$ could also contribute when one lepton results from the 
$W^\pm$ decay and the other lepton with same electric charge arises from 
leptonic B-meson decays. Among the above mentioned backgrounds, 
the dominant contribution in 2SSL channel comes from $WZ$, $ZZ$, and $t\overline{t}W$. 
In the 3SSL channel, following SM backgrounds are considered: 
$t\bar t W,~t\bar t Z,~t\bar t h~, WWZ,~WZZ,~ZZZ,~t\bar t t\bar t,$ and 
~$t\bar t b\bar b $. 
Out of all these possibilities, major contributions come from 
$t\overline{t}W$ and $WZZ$. All background events are generated using MadGraph and the 
cross-sections are taken upto NLO (see Refs.~\cite{Campbell:1999ah,Campanario:2008yg,Garzelli:2012bn,Heinemeyer:2013tqa,Yong-Bai:2015xna,Nhung:2013jta,Bredenstein:2009aj,Bevilacqua:2012em}).

%==========================================================
\subsection{Event Selection}
%===========================================================

%=============================================
\begin{figure}[!ht]
\centering
\includegraphics[width=7.7cm]{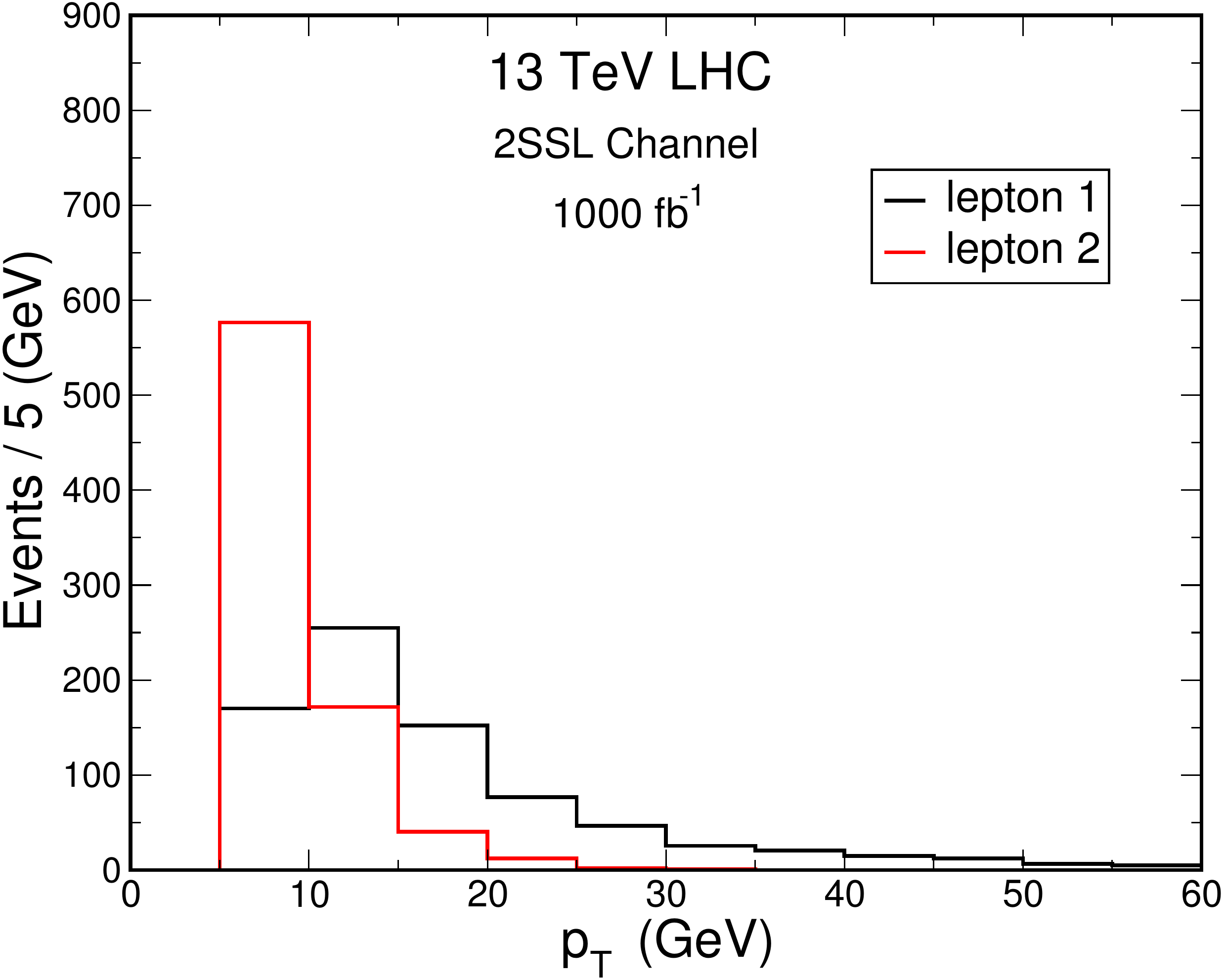}
\includegraphics[width=7.7cm]{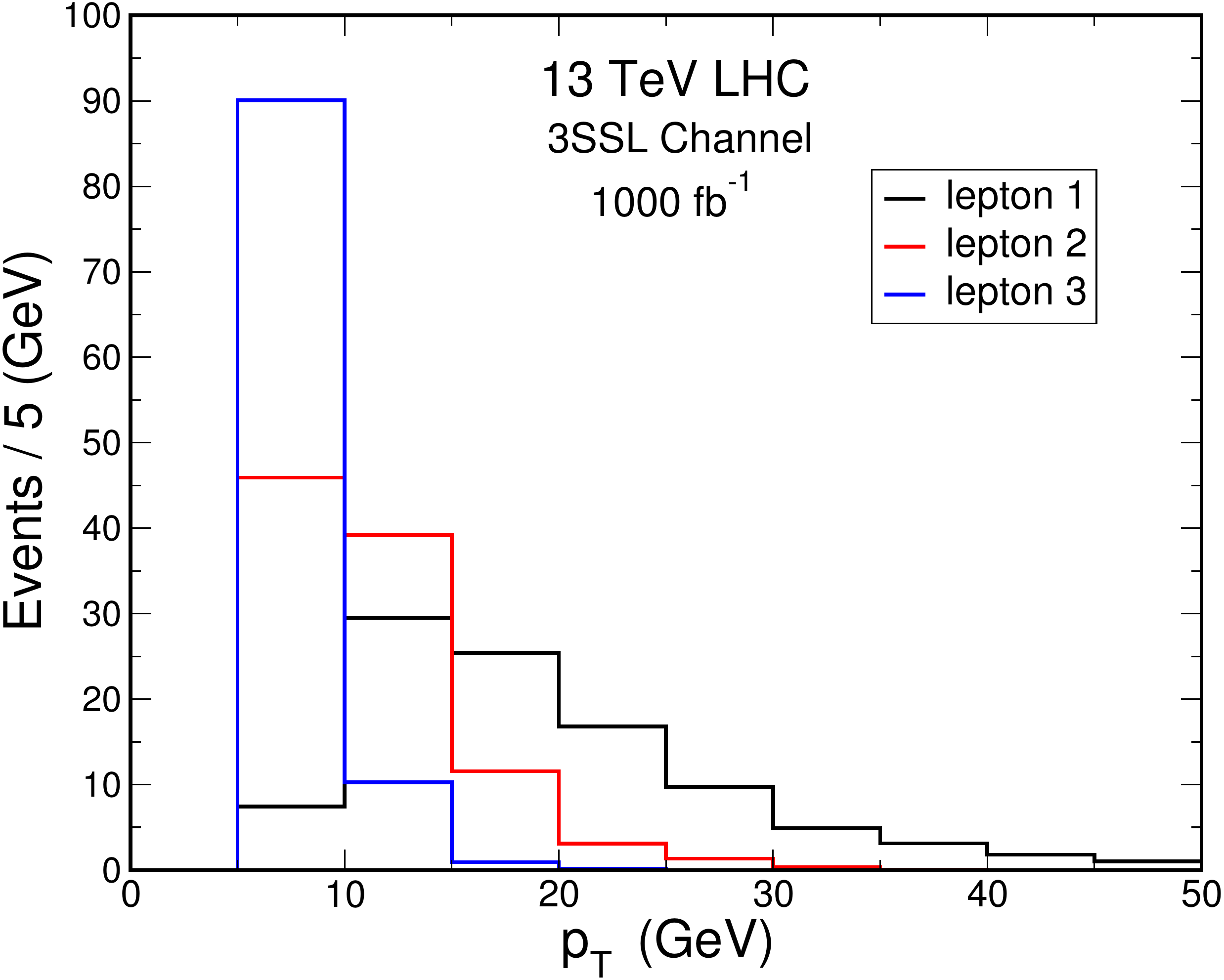}
\mycaption{The left panel shows the distributions of the leading and subleading lepton transverse momentum ($p_T(l)$) for signal events in the 2SSL channel 
at the 13 TeV LHC. The right panel depicts the same in the 3SSL channel. Here leptons are $l = e, \mu$. In both the panels, the events are weighted at 1000 fb$^{-1}$.}
\label{fig:pt}
\end{figure} 
%==============================================

%=============================================
\begin{figure}[!ht]
\centering
\includegraphics[width=7.7cm]{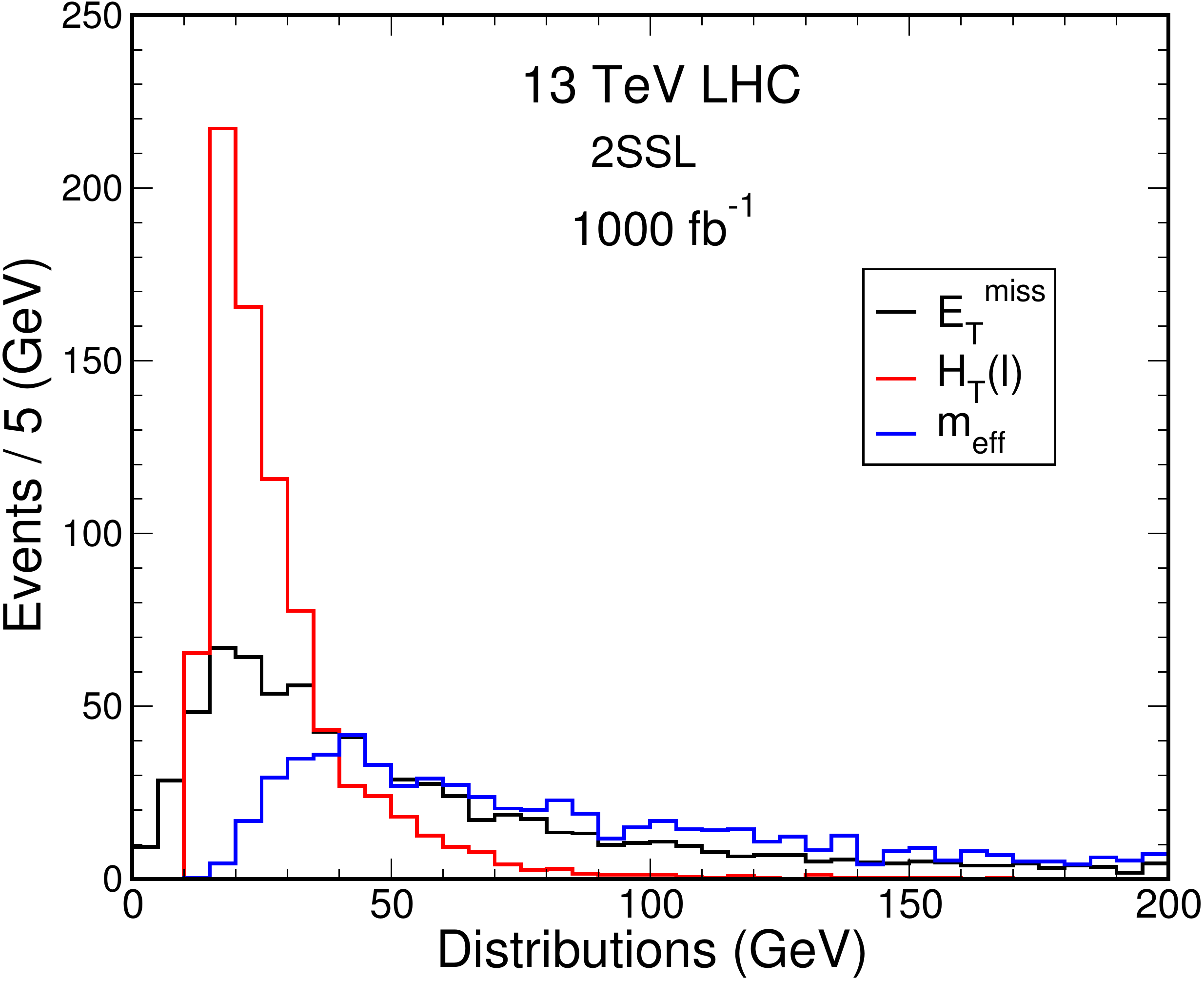}
\includegraphics[width=7.7cm]{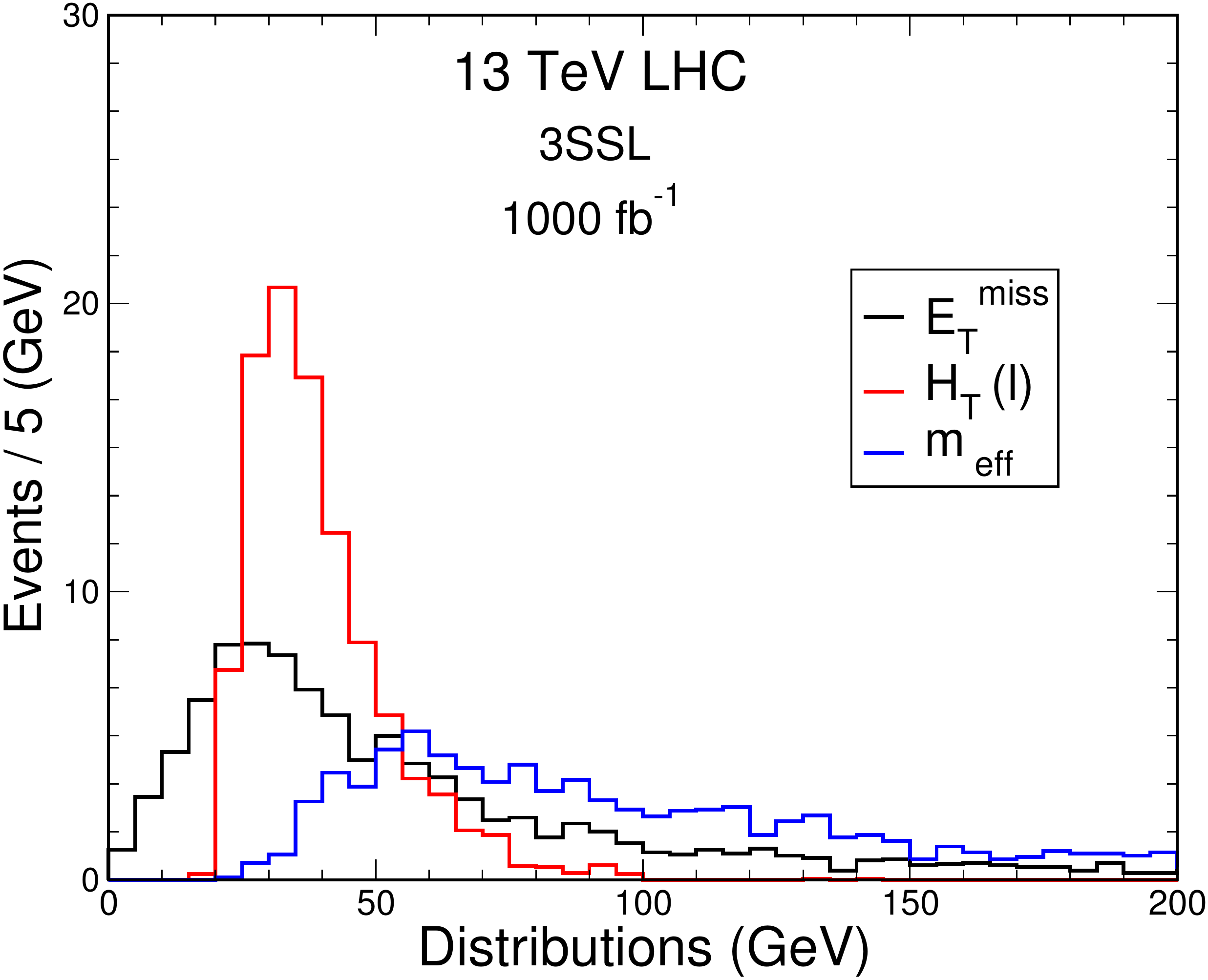}
\mycaption{The left panel shows the distributions of the missing energy ($\missET$), 
sum of the lepton transverse momentum ($H_T(l)= \sum_i p_T(l_i)$), and effective mass ($\meff= E_T+ H_T(l)+ H_T(j)$) for signal events in the 2SSL channel 
at the 13 TeV LHC. The right panel depicts the same in the 3SSL channel. Here leptons are $l = e, \mu$. In both the panels, the events are weighted at 1000 fb$^{-1}$.}
\label{fig:meff}
\end{figure} 
%==============================================

%===========================================
\begin{figure}[!ht]
\centering
\includegraphics[width=7.7cm]{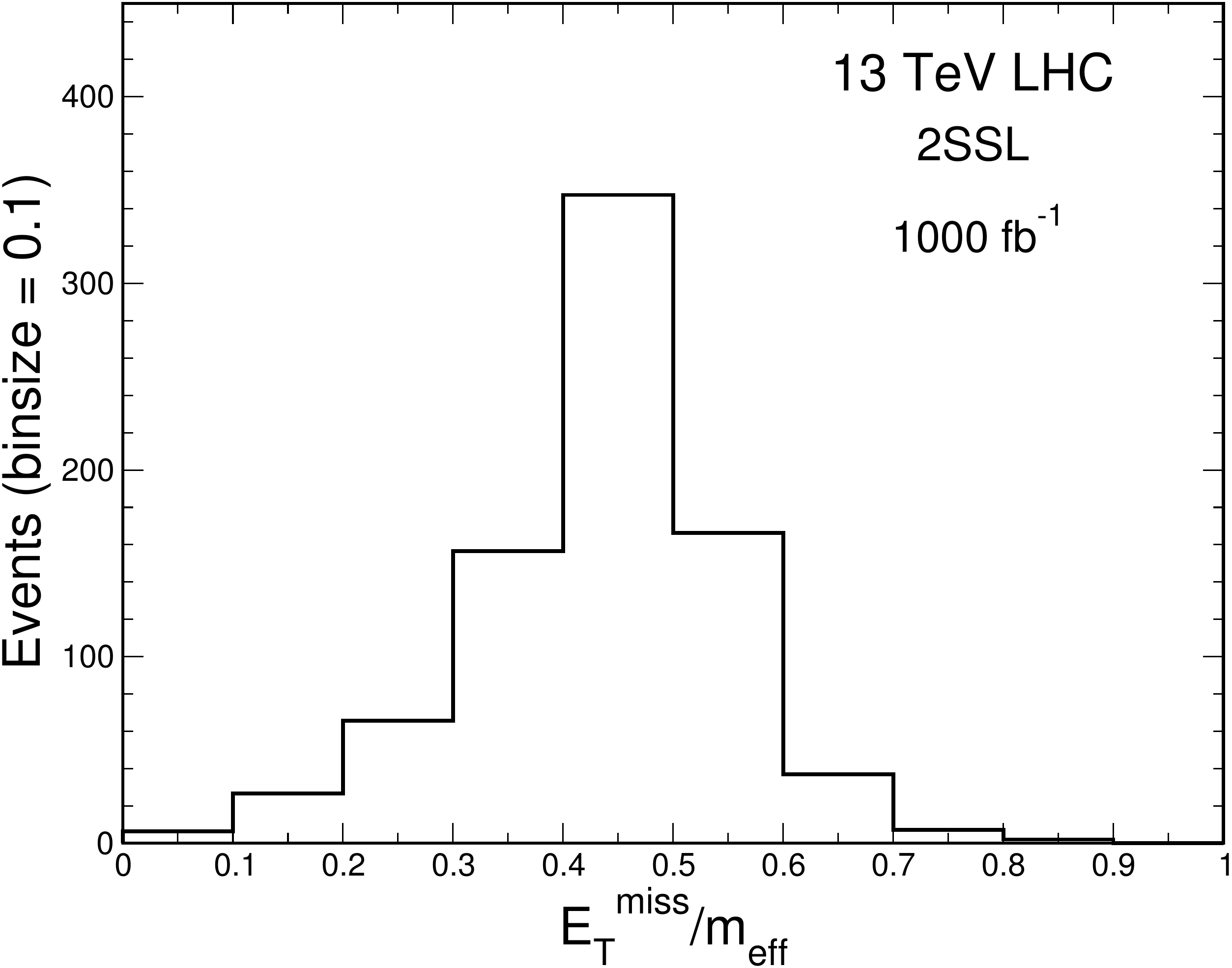}
\includegraphics[width=7.7cm]{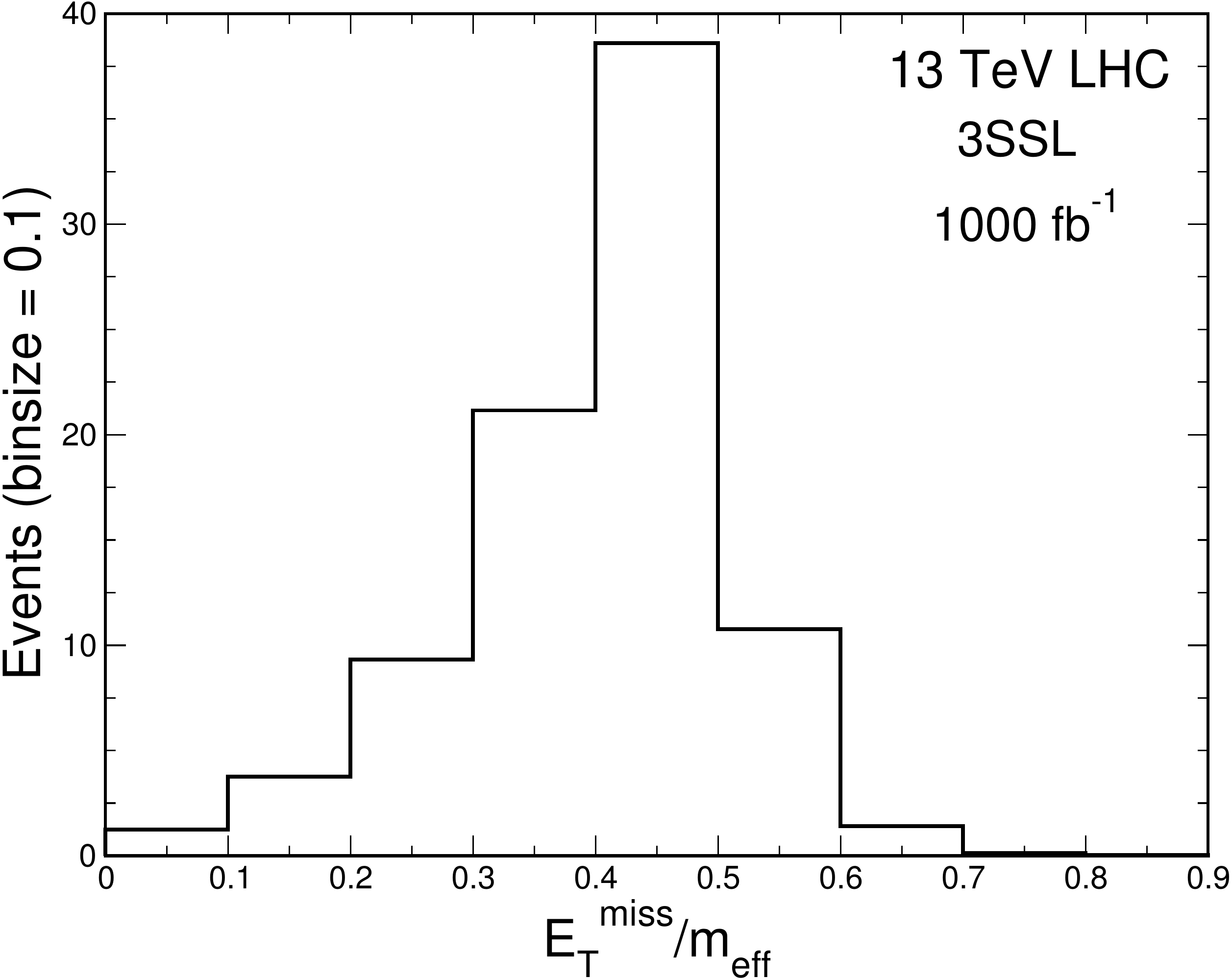}
\mycaption{The distribution of the ratio of the missing energy and 
the effective mass ($\missET/\meff$) for signal events in the 2SSL (left panel) 
and 3SSL (right panel) channels at the 13 TeV LHC. The events are weighted for 1000 fb$^{-1}$.}
\label{fig:et_meff}
\end{figure} 
%==============================================

Following are the generator level acceptance cuts that we impose while simulating the 
signal and backgrounds.
%=========
\begin{itemize}
\item For any lepton: $p_{T}\,(e,\mu) > 6$ GeV.
\item For any jet: $p_{T}\,(j) > 20$ GeV.
\item $|\eta (e)| < 2.5$, $|\eta (\mu)| < 2.5$, and  $|\eta (j)| < 2.4$.
\item $\Delta R_{l,l'} > 0.2$  ($l,l'=e,\mu$), $\Delta R_{j,j} > 0.5$, and $\Delta R_{\ell,j} > 0.4$.
\end{itemize}
%========
The transverse momentum distributions for the leading and subleading signal leptons 
are given in Fig.~\ref{fig:pt} (left panel for 2SSL and right panel for 3SSL). It is evident from 
this figure that 
the signal leptons are soft, as discussed earlier. Since most of the 
signal events are distributed in the lower $p_T$ region, a cut on 
the upper limit of $p_T(l)$ will effectively reduce the background. The other kinematic variables considered 
in the study are the sum of the transverse momentum of the leptons and jets, and the effective mass which are 
defined as: 
\begin{equation}
H_T(l)= \sum_{i} p_T(l)_i,~~~~
H_T(j)= \sum_{i} p_T(j)_i,~~~~
\meff= E_T+ H_T(l)+ H_T(j).\\
\end{equation}
The distributions of the $\missET$, $H_T(l)$, and $\meff$ are shown 
in Fig.~\ref{fig:meff} for the 2SSL (left panel) and 3SSL (right panel) channel. 
We find that an upper limit on $H_T(l)$ and $H_T(j)$ help to reduce 
the backgrounds significantly. In Fig.~\ref{fig:et_meff}, we plot distribution of the $\missET/\meff$ 
for both the channels. 
In the next section, we find that a cut on the lower values of the $\missET/\meff$ turns out to be quite effective 
to reduce the QCD-jet backgrounds. The backgrounds from $t\overline{t}W$ and $t\overline{t}Z$ can be reduced 
by applying a b-jet veto. Another effective way to reduce the same-sign multilepton backgrounds is to 
consider only small values of the transverse mass ($m_T$).
We imposed further selection criteria on these kinematic variables 
to enhance the signal to background ratios in the 2SSL, 3SSL, and 4SSL channels. 
The final event selection criteria are listed in Table~\ref{tab:a} and the effects are
discussed in the next section.

%================================
\begin{table}[ht!]
\centering
\begin{tabular}{|c|c|c|c|}
\hline\hline
Selection Cuts & 2SSL &  3SSL& 4SSL \\
\hline\hline
$S1$ & $l^+l^+$ or $l^-l^-$ & $l^+l^+l^+$ or $l^-l^-l^-$ & $l^+l^+l^+l^+$ or $l^-l^-l^-l^-$\\
\hline\hline
$S2$ &  $b-jets = 0$             &  $b-jets = 0$                  &  $b-jets = 0$ \\
                 & $p_T(l_1)< 35$ GeV        & $p_T(l_1)< 35$ GeV             & $p_T(l_1)< 35$ GeV\\
                 & $p_T(l_2) < 20$ GeV      & $p_T(l_2) < 20$ GeV            & $p_T(l_2) < 20$ GeV\\
                 & $\missET/\meff > 0.2$    & $p_T(l_3) < 15$ GeV             & $p_T(l_3) < 15$ GeV\\
                 & $H_T(l) < 60$ GeV            & $\missET/\meff > 0.2$            & $p_T(l_4) < 15$ GeV\\
                 & $H_T(j) < 200$ GeV            & $H_T(l) < 60$ GeV                    & $\missET/\meff > 0.2$\\
                  &                           &$H_T(j) < 200$ GeV                    & $H_T(l) < 60$ GeV\\
                 &                            &                                  & $H_T(j) < 200$ GeV\\
\hline\hline
$S3_{x}$ & $m_{T} < 40$ GeV &  - &- \\
                 & $\missET > 10$ GeV    & - &- \\
\hline\hline                 
$S3_{y}$  & $m_{T} < 40$ GeV &  - &-\\
                 & $\missET > 20$ GeV    & - &- \\
\hline\hline
$S3_{z}$  & $m_{T} < 40$ GeV &  - &- \\
                 & $\missET > 30$ GeV    & - &- \\
\hline\hline
\end{tabular}
\mycaption{Selection criteria ($S1$, $S2$, and $S3$) that we consider for the 2SSL, 3SSL, and 4SSL channels. In selection 
$S1$, we only consider the events with exactly 2, 3 or 4 leptons with same sign. 
Other events containing opposite sign leptons are vetoed. Here $l=e, \mu$. 
Also, note that for the 3SSL and 4SSL channels, we do not impose selection criteria $S3$. }
\label{tab:a}
\end{table}
%====================================
 
%==========================================
\subsection{Exclusion Limits and Discovery Reach in the Same-sign Multilepton Channels}
%==========================================
As discussed in Table~\ref{tab:a}, the same-sign 
lepton ($l=e,\mu$) final states are required to pass 
the selections $S1$, $S2$, and $S3$ in the 2SSL channel. For 3SSL, and 4SSL channels we 
only impose $S1$ and $S2$. 
While analyzing the same-sign multilepton channels, we find that the remaining background is 
small after passing through several selections.
If the background is small then the standard formula ($s/\sqrt{b}$)
overestimates the discovery significance. Also, in real experiments, the background ($b$) is never known with 100\% accuracy. 
Therefore, while calculating the discovery significance and exclusion, 
we include an uncertainty in the background ($\Delta_b$)~\cite{Cowan:SLAC2012}. This analysis is helpful to deal with small backgrounds, implemented before in Ref.~\cite{Kumar:2015tna}. The significance for discovery is defined as, 
\begin{equation}
\Zdis = \left [ 2\left ((s+b) \ln \left [\frac{(s+b)(b+\sigmab^2)}{b^2+(s+b)\sigmab^2}\right ]-\frac {b^2}{\sigmab^2} \ln\left [1+ \frac {\sigmab^2 s}{b(b+\sigmab^2)}\right ]\right)\right]^{1/2} .
\label{eq:zdis}
\end{equation}
If $\Delta_b = 0$,
\begin{equation}
\Zdis=\sqrt{2[(s+b)\ln(1+s/b)-s]}.
\label{eq:zdis1}
\end{equation}
In the above equation, if $b$ is large, then we obtain the well known expression
\begin{equation}
\Zdis = s/\sqrt{b}.
\label{eq:zdis2}
\end{equation}
It is evident from the above discussion that if $b$ is small, $s/\sqrt{b}$ overestimates the significance. Therefore, we use the expression given in Eq.~\ref{eq:zdis} to estimate the discovery reach by assuming $\Zdis \geq 5$ which corresponds to 5$\sigma$ discovery ($p<2.86 \times 10^{-7}$). To set the exclusion limit at a given confidence level (CL), we use the following expression
\begin{equation}
\Zexc=\left [2 \left \{ s-b \ln \left (\frac{b+s+x}{2b} \right ) - \frac{b^2}{\Delta_b^2} \ln \left (\frac{b-s+x}{2b} \right ) \right \} -
(b + s - x) (1 + b/\Delta_b^2) \right ]^{1/2},
\label{eq:zexc}
\end{equation}
where
\begin{equation}
x = \sqrt{(s+b)^2 - 4 s b \Delta_b^2/(b + \Delta_b^2)}.
\end{equation}
In the above equation, if $\Delta_b = 0$,
\begin{equation}
\Zexc = \sqrt{2(s - b \ln(1 + s/b))}.
\end{equation}
For a median expected 95\% CL exclusion
($p = 0.05$), we use $\Zexc \geq 1.645$ for different 
$\Delta_b $.
With increasing amount of LHC data, and hence for a better understanding of the detector response, 
the experimental uncertainties in the estimation of the SM backgrounds are expected to be reduced 
significantly. Therefore, we assume that the systematic uncertainty in background estimation falls as $1/\sqrt{\mathcal{L}}$ and we also incorporate this effect in $\Delta_b$. We choose different values of the systematic uncertainties at 10\%, 25\% and 50\% at 10 $fb^{-1}$ integrated luminosity and scale the uncertainty appropriately for higher luminosities. 

%=======================
% Discussion on 2SSL Channel
%=======================

%===========================
\begin{table}[ht!]
\centering
\begin{tabular}{|c|c|c|c|c|c|c|}
\hline\hline
$M_{\chi^{n\pm}}$ (GeV) & $\sigma (fb)\times$ BR & $S1$ (fb) & $S2$ (fb) & $S3_{x}$(fb) & $S3_{y}$ (fb) & $S3_{z}$ (fb)\\
\hline\hline
200,205,216,233,257 & 214.4& 2.15&1.88&1.26&0.82&0.48 \\
300,307,322,345,377 & 75.8 & 2.22&1.85&1.12&0.77&0.44 \\
400,409,428,456,494 & 24.7& 1.50&1.44&0.58&0.40&0.22 \\
500,511,533,566,610 & 9.6 & 0.96&0.61&0.26&0.17&0.09 \\
600,613,638,675,725 & 4.5 & 0.61&0.32&0.12&0.08&0.05 \\
700,714,742,783,838 & 2.3 & 0.38 & 0.17 & 0.05 & 0.04 & 0.02 \\
800,816,846,891,950 & 1.27 & 0.25 & 0.10& 0.03 &0.02 & 0.01 \\
\hline\hline
\end{tabular}
\mycaption{Signal cross-sections after passing the various selection criteria in the 2SSL channel at the 13 TeV LHC for different masses of the quintuplet. Here $n$ runs from 0 to 4 in a given quintuplet. We assume BR($\chi^{Q+1} \rightarrow l \nu \chi^{Q}$) = 0.22 where $l=e,\mu$ and BR($\chi^{Q+1} \rightarrow q \overline{q} \chi^{Q}$) = 0.67.}
\label{tab:3}
\end{table}
%============================

%============================
\begin{table}[ht!]
\centering
\begin{tabular}{|c|c|c|c|c|c|c|}
\hline\hline
Process & Cross-section (fb) & $S1$ (fb)& $S2$ (fb)& $S3_{x}$  (fb) & $S3_{y}$  (fb)& $S3_{z}$ (fb) \\
\hline\hline
$p p \rightarrow WW$ & 130$\times 10^{3}$ 	& 0.234    & 0.117 	& $<10^{-4}$  & $<10^{-4}$  & $<10^{-4}$  \\
$p p \rightarrow WZ$ & 49$\times 10^{3}$ 	 & 98.365    & 14.56 	& 3.84       & 2.196         & 1.124  \\
$p p \rightarrow ZZ$ 	& 16$\times 10^{3}$ 	& 7.303     & 0.654 	& 0.286       & 0.182        & 0.103  \\
$p p \rightarrow hh$ & 40$\times 10^{3}$  	& 0.083     & 0.012  	& 0.004      & 0.004         & 0.003 \\
$p p \rightarrow hZ$ & 0.97$\times 10^{3}$  	 & 0.890     & 0.122  	& 0.046      & 0.031         & 0.015 \\
$p p \rightarrow t\overline{t}W$ & 600                   & 1.574     & 0.014     &0.002        & 0.002        & 0.002 \\
$p p \rightarrow t\overline{t}H$ & 508                   & 0.503     & 0.004     & $< 10^{-4}$     & $<10^{-4}$     & $ <10^{-4}$ \\
$p p \rightarrow t\overline{t}Z$ & 920                   &  0.468     & 0.002    & 0.0002        & 0.0002             & 0.0002 \\
$p p \rightarrow t\overline{t}$ & 820$\times 10^{3}$     & 0.369  & 0.025     & $< 10^{-4}$    & $<10^{-4}$      & $<10^{-4}$ \\
\hline\hline
Total & - & 109.79 & 15.515 & 4.176 & 2.416 &1.247 \\
\hline\hline
\end{tabular}
\mycaption{Background cross-sections in the 2SSL channel after passing various selection criteria at the 13 TeV LHC.}
\label{tab:4}
\end{table}
%==============================
For the two same-sign lepton channel (2SSL), we require exactly 2 leptons ($ee$ or $\mu \mu$ or $e \mu$). 
The selection criteria in 2SSL channel are listed in Table~\ref{tab:a}.
The signal cross-section after passing various selections ($S1$, $S2$, and $S3$) are given in Table~\ref{tab:3} for 
2SSL channel at the 13 TeV LHC. The signal events suffer from a very low missing transverse energy ($\missET$), which comes mainly 
from the neutral state of the quintuplet ($\chi^0$). So, enforcing a cut on $\missET$ will yield smaller signal cross-section, as can be seen in 
Table~\ref{tab:3}. On the other hand, a cut on the lower limit of the missing energy effectively 
reduces the probability of jet faking leptons in the final state~\cite{Ozcan:2009qm}. 
So, here we choose three different $\missET$ cuts, $S3_x$, $S3_y$, and $S3_z$.
The background cross-section to pass the same selections are given in Table~\ref{tab:4} for 2SSL channel.
We find that the selection $S3$ is very effective in reducing the backgrounds.

%=============================================
\begin{figure}[!ht]
\centering
\includegraphics[width=7.7cm]{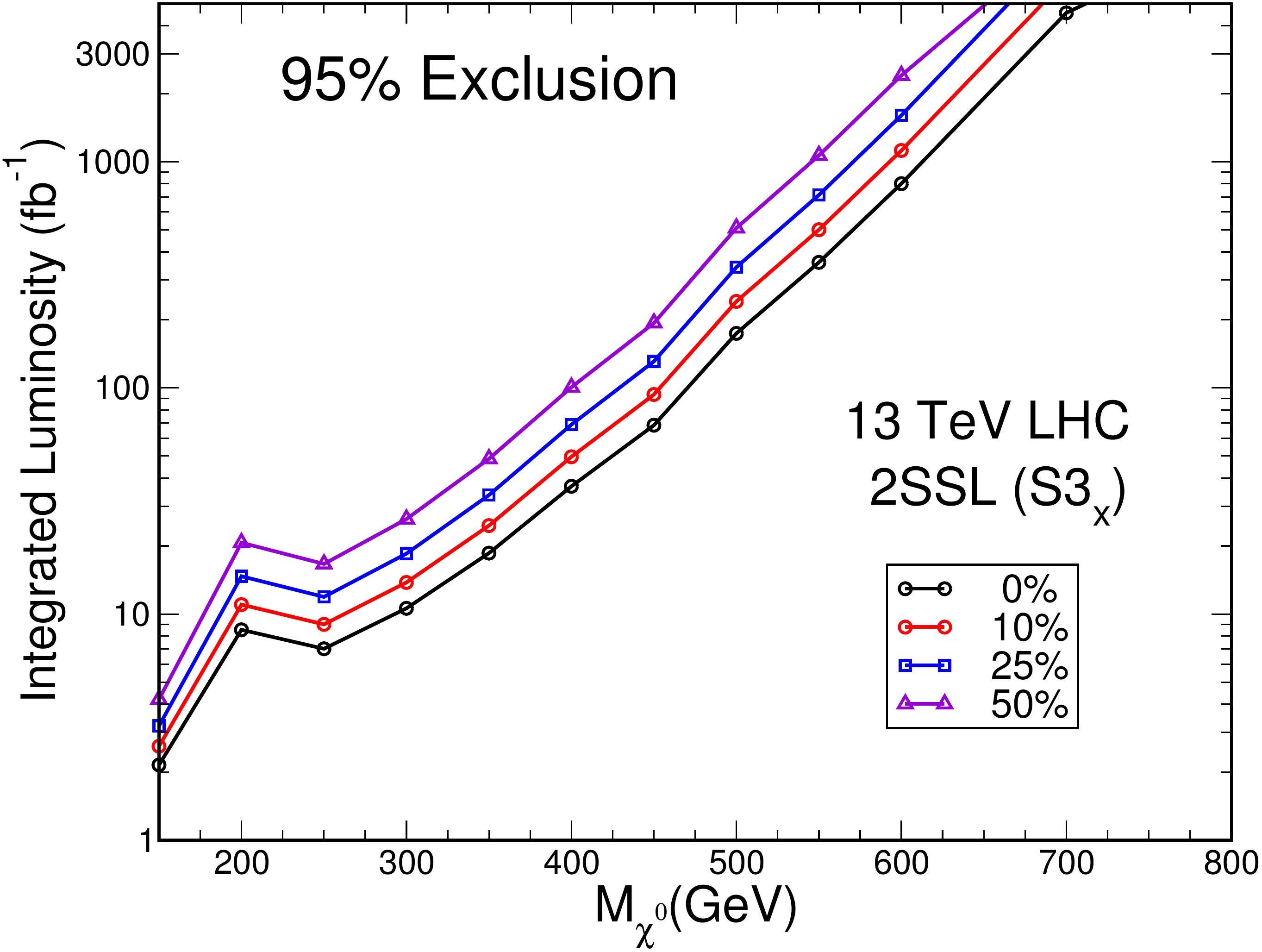}
\includegraphics[width=7.7cm]{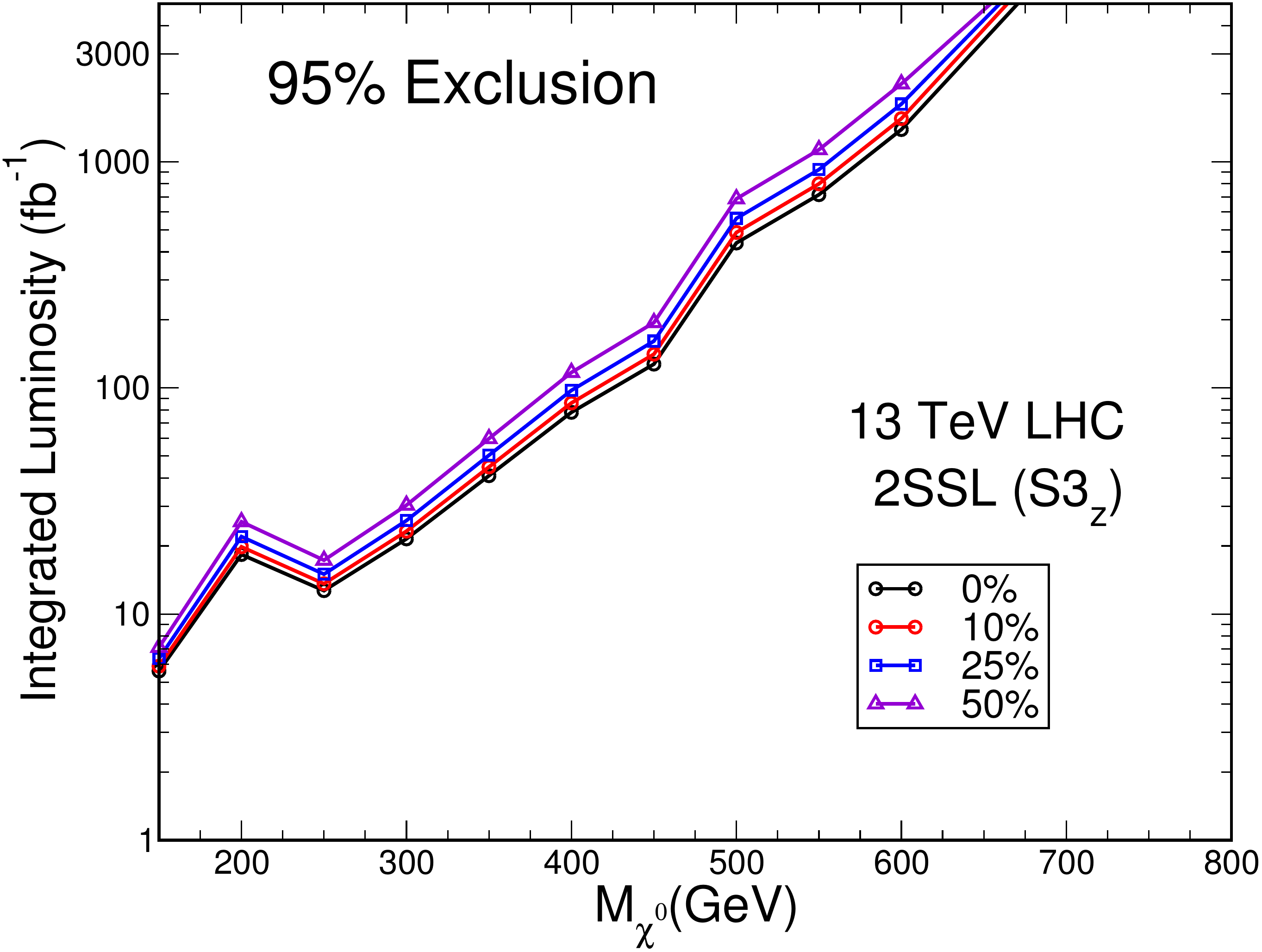}
\mycaption{In the left panel, we show the required integrated luminosity for 95\% CL exclusion ($Z_{\text{exc}}\geq$ 1.645) after imposing the $S3_x$ selection criteria in the 2SSL channel at the 13 TeV LHC as a function of $M_{\chi^0}$. The right panel depicts the same for selection criteria $S3_z$. The colored lines in both the panels are for the different level of uncertainties in the background events ranging from 0 to 50$\%$.}
\label{result2sslexc}
\end{figure} 
%==============================================

%====================================================
\begin{figure}[!ht]
\centering
\includegraphics[width=7.7cm]{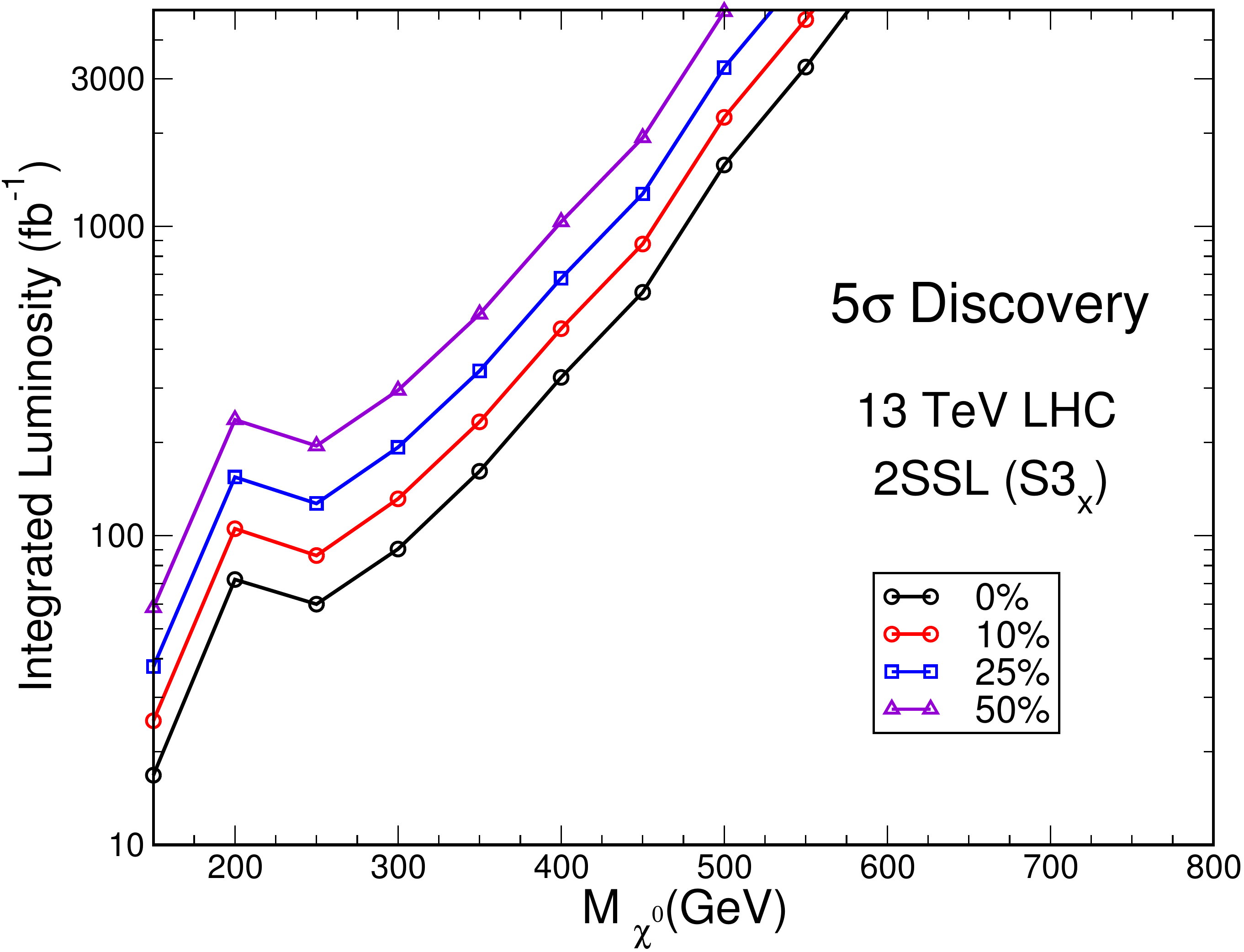}
\includegraphics[width=7.7cm]{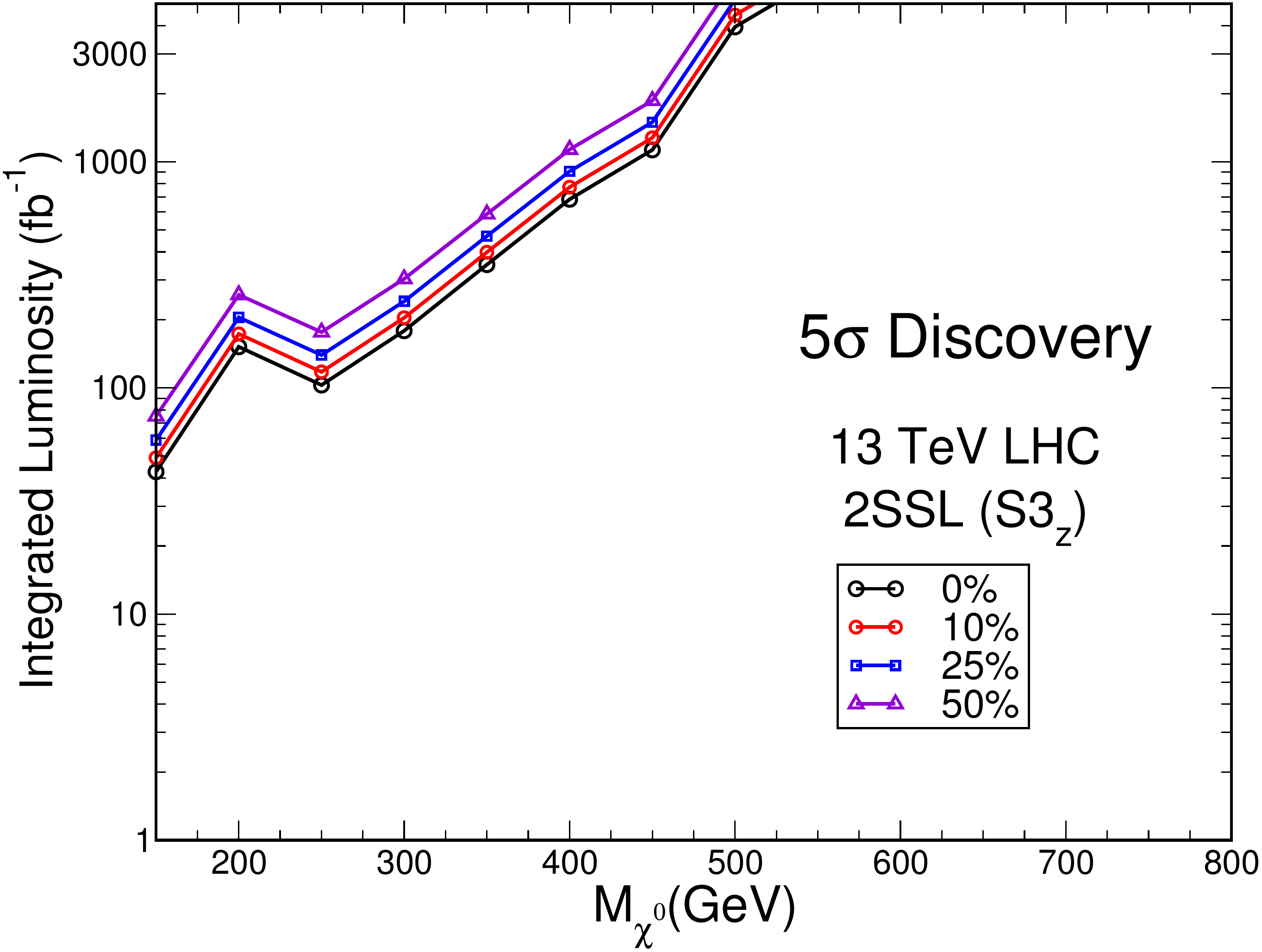}
\mycaption{In the left panel, we show the required integrated luminosity for 5$\sigma$ discovery ($Z_{\text{dis}} \geq$ 5) after imposing the $S3_x$ selection criteria in the 2SSL channel at the 13 TeV LHC as a function of $M_{\chi^0}$. The right panel depicts the same for selection criteria $S3_z$. The colored lines in both the panels are for the different level of uncertainties in the background events ranging from 0 to 50\%.}
\label{result2ssldis} 
\end{figure} 
%=====================================================

In Fig.~\ref{result2sslexc} and Fig.~\ref{result2ssldis}, 
projections of the required luminosity for $95\%$ confidence level exclusion 
($Z_{exc}\geq1.645$) and $5\sigma$ discovery ($Z_{dic}\geq5$) 
are shown as a function of the neutral state mass ($M_{\chi^0}$), 
for two different selections $S3_{x}$ and $S3_{z}$. To make the predictions we use Eq.~\ref{eq:zdis} and Eq.~\ref{eq:zexc} and vary the uncertainty 
in the backgrounds between $0-50\%$. As a larger $\missET$ 
cut further reduces the signal cross-section, more luminosity is expected to be required to set an exclusion in $S3_{z}$ compared to 
$S3_{x}$. But from Fig.~\ref{result2sslexc}, it is clear that the cut on $\missET$ is not that sensitive 
since a strong $\missET$ cut also reduces the background cross-section at a comparable rate, 
as reflected in Table~\ref{tab:3} and Table~\ref{tab:4}. With 3000 $fb^{-1}$ integrated luminosity, one can exclude up to 610 GeV of the neutral 
state mass ($M_{\chi^0}$) if selection $S3_x$ is applied. 
Also from Fig.~\ref{result2ssldis}, we can say that after selection $S3_x$, even with 3000 $fb^{-1}$ integrated luminosity, the discovery prospects 
will be challenging if the neutral state mass is greater than 475 GeV.

%=======================
% Discussion on 3SSL Channel
%=======================

\begin{figure}[!ht]
\centering
\includegraphics[width=7.7cm]{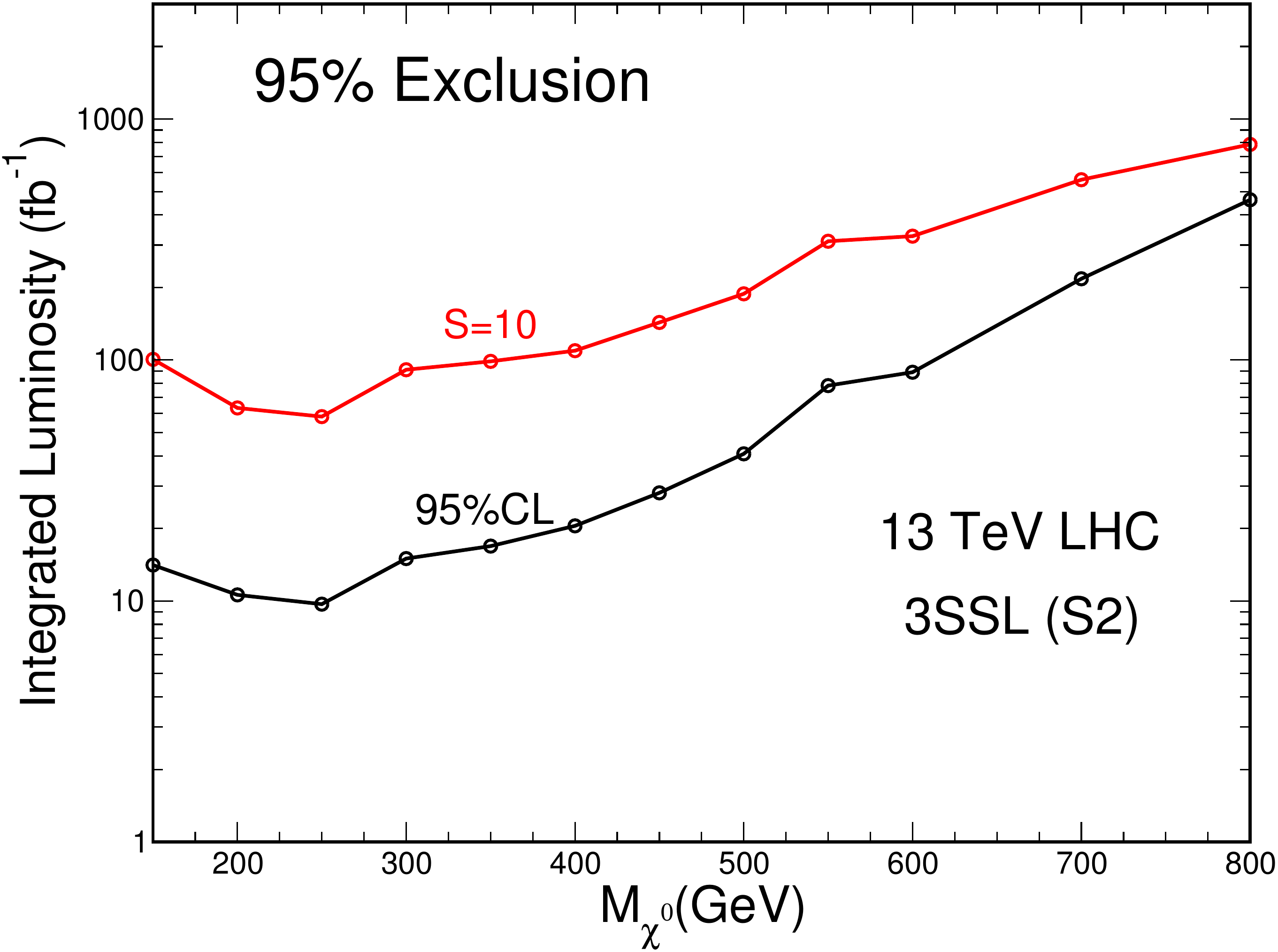}
\includegraphics[width=7.7cm]{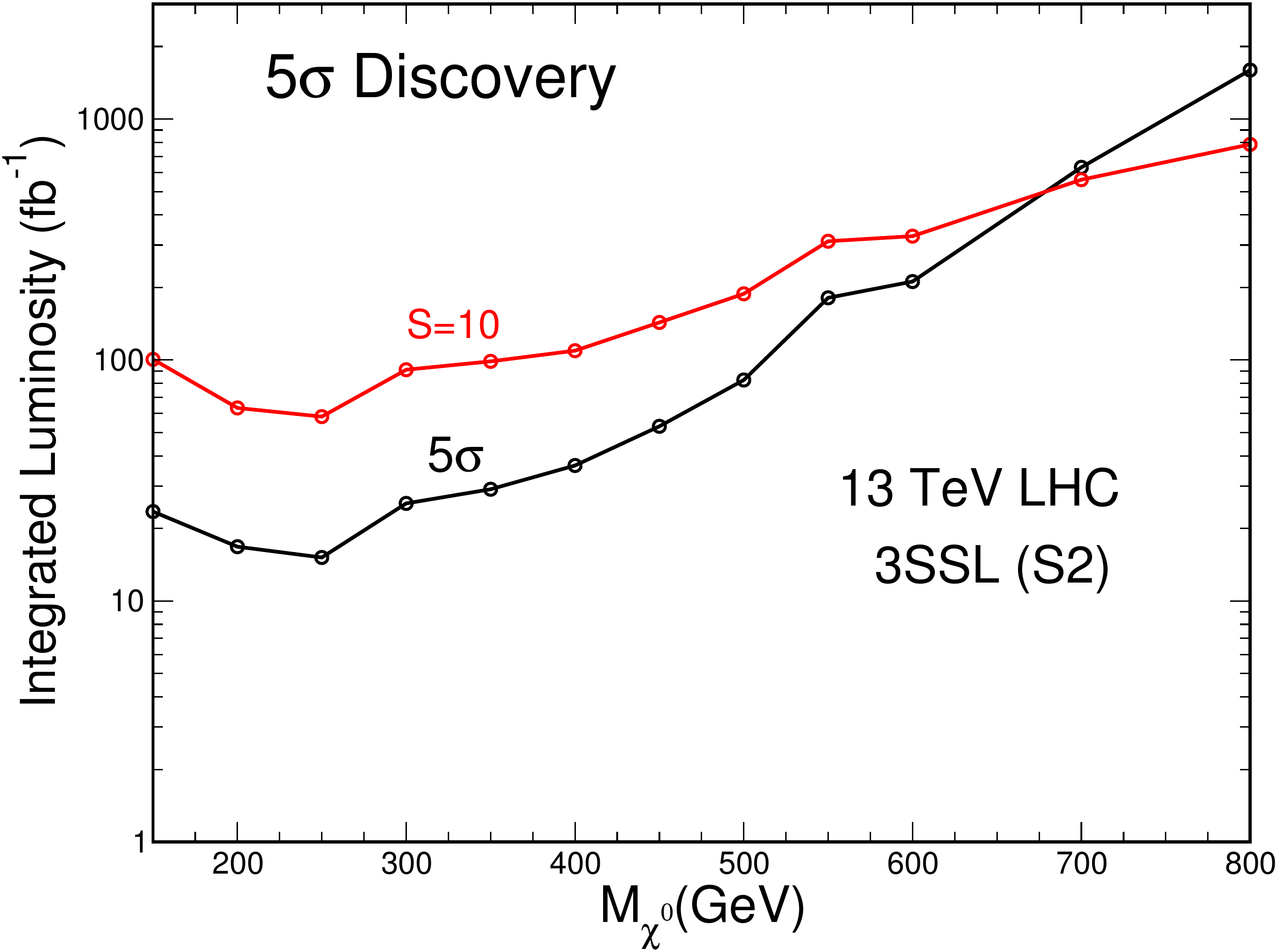}
\mycaption{The left panel shows the required integrated luminosity at the 13 TeV LHC for 95\% CL exclusion ($Z_{\text{exc}} \geq$ 1.645) in the 3SSL channel after imposing $S2$ selection criteria as a function of $M_{\chi^0}$. The right panel portrays the same for 5$\sigma$ discovery ($Z_{\text{dis}} \geq$ 5). Note that the results remain same for any uncertainty in the background in the range 0 to 50\%. In both the panels (red lines), we also show the required integrated luminosity to observe 10 signal events in the 3SSL channel.}
\label{result3ssl} 
\end{figure}

In the 3SSL channel, exactly three same-sign leptons
($eee$ or $\mu\mu\mu$ or $e\mu\mu$ or $ee\mu$ ) 
are required to pass the selections $S1$ and $S2$ (see Table~\ref{tab:a}). These two selections are 
sufficient to suppress the backgrounds. Moreover, as the signal cross-section is small, 
implementing the $S3$ selection criterion reduces the signal cross-section significantly. Table~\ref{tab:5} and Table~\ref{tab:6} gives 
the signal and background cross-section respectively after the selections ($S1$ and $S2$) for 3SSL channel at the 13 TeV LHC. 
It is clear from the table that selection $S2$ is sufficient to make prediction 
for exclusion and discovery as the total background becomes sufficiently small with $t\overline{t}W$ being the dominant 
background. The results for the exclusion and discovery are calculated as before and shown 
in Fig.~\ref{result3ssl}. A quintuplet with neutral state mass ($M_{\chi^0}$) 
up to 800 GeV can be excluded 
with only 500 $fb^{-1}$ luminosity. 
As the signal cross-section 
is small in 3SSL channel, we also plot the required luminosity to observe 10 signal 
events (red line in Fig.~\ref{result3ssl}). A quintuplet with neutral state mass $M_{\chi^0}\leq$ 800 GeV 
can be excluded with 800 $fb^{-1}$ luminosity if at least 10 signal events are required. 
It might be also possible to discover a $M_{\chi^0}$
up to 750 GeV with 1000 $fb^{-1}$ luminosity. 

%================================================
\begin{table}[ht!]
\centering
\begin{tabular}{|c|c|c|c|}
\hline\hline
$M_{\chi^{n\pm}}$ (GeV) & $\sigma (fb)\times$ BR & $S1$ (fb)& $S2$ (fb)\\
\hline\hline
200,205,216,233,257 & 71.9 & 0.158 & 0.129 \\
300,307,322,345,377 & 11.23 & 0.110&0.091 \\
400,409,428,456,494 & 3.7 & 0.091&0.067 \\
500,511,533,566,610 & 1.47& 0.053 & 0.034 \\
600,613,638,675,725 & 0.7 & 0.030 & 0.016 \\
700,714,742,783,838 & 0.35 & 0.017 & 0.007 \\
800,816,846,891,950 & 0.19 & 0.013 & 0.003 \\
\hline\hline
\end{tabular}
\mycaption{Signal cross-sections after passing the various selection criteria in the 3SSL channel at the 13 TeV LHC for different masses of the quintuplet. Here $n$ runs from 0 to 4 in a given quintuplet. We consider BR($\chi^{Q+1} \rightarrow l \nu \chi^{Q}$) = 0.22 where $l=e,\mu$ and BR($\chi^{Q+1} \rightarrow q \overline{q} \chi^{Q}$) = 0.67.}
\label{tab:5}
\end{table}
%====================================================

%================================================
\begin{table}[ht!]
\centering
\begin{tabular}{|c|c|c|c|}
\hline\hline
Process & Cross-section (fb) &  $S1$ (fb)& $S2$ (fb) \\
\hline\hline
$p p \rightarrow t\overline{t}h$ & 508.5 & 9.1$\times 10^{-4}$ & 0 \\
$p p \rightarrow t\overline{t}Z$  & 920 & 3.6$\times 10^{-4}$ & 0 \\
$p p \rightarrow t\overline{t}W$  & 600 & 2.9$\times 10^{-4}$ & 1.5$\times 10^{-4}$ \\
$p p \rightarrow WWZ$ &103 & 0 & 0 \\
$p p \rightarrow WZZ$ &66 & 1.5$\times 10^{-3}$ & 8.5$\times 10^{-6}$ \\
$p p \rightarrow ZZZ$ &13.3 & 1.5$\times 10^{-4}$ & 0 \\
$p p \rightarrow t\overline{t}t\overline{t}$  & 15.33 & 0.91$\times 10^{-4}$&0 \\
$p p \rightarrow t\overline{t}b\overline{b}$  & 2638 & 2.2$\times 10^{-4}$ &0 \\
\hline\hline
Total & -& $3 \times 10^{-3}$ & $1.55 \times 10^{-4}$ \\
\hline\hline
\end{tabular}
\mycaption{Background cross-sections in the 3SSL channel after passing the 
selection criteria $S1$ (third column) and $S2$ (fourth column) at the 13 TeV LHC.}
\label{tab:6}
\end{table}
%===============================================

%====================================================
%=====================================================

%=======================
% Discussion on 4SSL Channel
%=======================

%=============================================
\begin{figure}[!htb]
\centering
\includegraphics[width=0.5\textwidth]{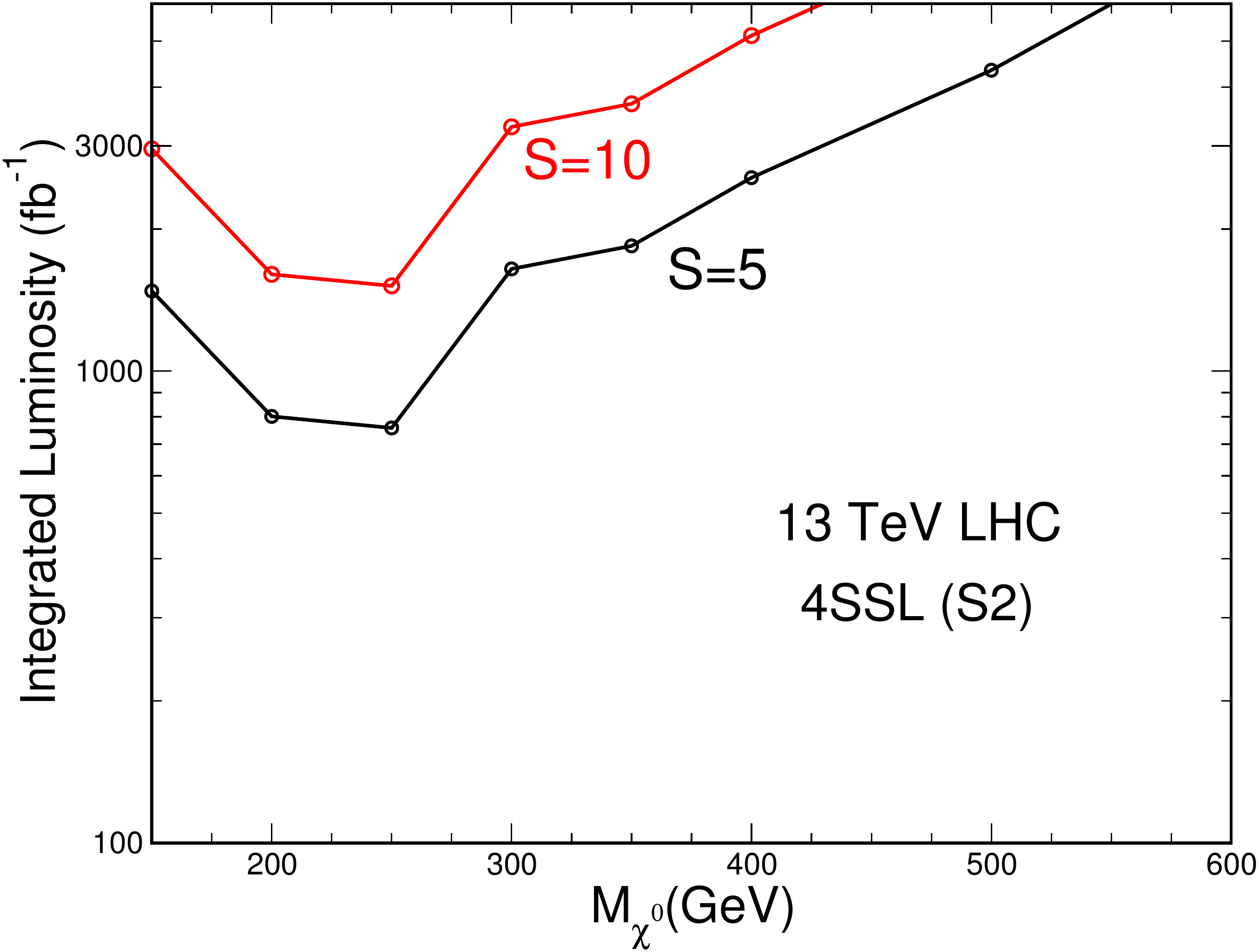}
\mycaption{The black and red lines depict the required 
integrated luminosity at the 13 TeV LHC to observe at least 
5 and 10 signal events respectively in the 4SSL channel 
after imposing the $S2$ selection criteria as a function 
of the neutral state mass ($M_{\chi^0}$).}
\label{result4ssl}
\end{figure} 
%==============================================

The 4SSL channel can provide 
a very clean signature as the background is almost nonexistent. 
The events are required to pass both $S1$ and $S2$, but 
the signal cross-section becomes quite small if we impose $S2$. 
For example, if the mass of the neutral state of the 
quintuplet is 300 GeV, the signal cross-section after passing 
$S2$ is only 0.003 fb. If the same selection criteria is applied, 
this channel has practically zero background, smaller by a factor of $\alpha$ compared to 3SSL channel. 
Hence, in Fig.~\ref{result4ssl}, we only plot the required 
luminosity to observe 5 and 10 signal events as 
a function of the neutral state mass ($M_{\chi^0}$) at 13 TeV if 
selection $S2$ is imposed. We find that for 300 GeV 
neutral state mass, the required luminosity will be greater than 3000 $fb^{-1}$ 
in order to see 10 signal events.

%======================
\section{Concluding Remarks}
%======================

In this paper, we consider the left-right symmetric model with a $SU(2)_R$ quintuplet fermion. The neutral state of this quintuplet, which is also the lightest, can be a viable dark matter candidate, while its charged fields can give rich collider signatures in the form of the multilepton final states. We study in detail the collider imprints of the same-sign multilepton final states in the context of the 13 TeV LHC. We consider the scenario where the quintuplet fermion possesses $(B - L) = 4$ charge, since it gives the most interesting collider signatures due to the presence of high charge-multiplicity particles producing lepton-rich final states.

These high charge-multiplicity particles can be pair produced through quark-antiquark initiated Drell-Yan processes or through photo production by photon-photon fusion process. Here, we show that the photon-photon fusion process contributes significantly in the total cross-section of the quintuplet fermions at the 13 TeV LHC. They can also have associated production into two quintuplet fermions of different charges through a $W_R$ boson. But, the production cross-section of this channel would be negligibly small. Once produced, the quintuplet fermions decay into the next lightest member of that quintuplet and two leptons or two quarks through an off-shell $W_R$ boson. This ultimately gives rise to final states comprising of a number of leptons, jets, and missing energy. Here, we study the same-sign multilepton signatures such as 2SSL, 3SSL, and 4SSL final states in the context of the LHC experiment at 13 TeV. 

For our analysis, we consider several benchmark points with the neutral state mass ($M_{\chi^0}$) of the quintuplet ranging from 200 GeV to 800 GeV. For the 2SSL case 
$M_{\chi^0} \leq 610$ GeV can be we can excluded at 95\% confidence level for 3000 $fb^{-1}$ luminosity. The discovery prospect at 5$\sigma$ level, being much more challenging, has an upper limit of $M_{\chi^0} \leq 475$ GeV. As far as the 3SSL channel is concerned, it has a much lower background as compared to the 2SSL channel. This enhances both the exclusion and discovery potential in the 3SSL channel. The LHC at 13 TeV would be able to exclude a $M_{\chi^0} \leq 800$ GeV at 95\% CL with an integrated luminosity of 500 $fb^{-1}$. Discovery at 5$\sigma$ is achievable upto $M_{\chi^0} \leq 750$ GeV with 1000 $fb^{-1}$ luminosity. The 4SSL is the cleanest channel with almost zero background, but at the same time it will have a very small signal cross-section which makes it extremely difficult to place a limit for exclusion or predict any discovery potential. 

Overall, our study reflects that with the 13 TeV LHC, a $5\sigma$ discovery reach or a 95\% CL exclusion limit on the $SU(2)_R$ quintuplet mass can be achieved by observing the same-sign multilepton signatures. We hope that the study performed in this paper will add to the long standing quest to search for new physics at LHC through the same-sign multilepton final states.

%==========================
\subsubsection*{Acknowledgments}
%==========================
We thank Dr. Biplob Bhattacherjee for useful comments.
S.K.A. acknowledges the support from the DST/INSPIRE 
Research Grant [IFA-PH-12], Department of Science and 
Technology, India and the Young Scientist Project 
[INSA/SP/YSP/144/2017/1578] from the Indian National Science Academy.
K.G. acknowledges the support from the DST/INSPIRE Research Grant 
[DST/INSPIRE/04/2014/002158]. N.K. acknowledges the support from the Indo-French 
Center for Promotion of Advanced Research (CEFIPRA Project No. 5404-2). 
N.K. would like to thank Institute of Physics, Bhubaneswar for hospitality at the initial stage of the project.
A.P. is supported by the SERB National Postdoctoral fellowship [PDF/2016/000202]. 
A.P. and N.K. thank the organizers of ``Candles of Darkness" 
conference held at ICTS, Bangalore, India during June, 2017 for giving 
the opportunity to present the preliminary results of this work. 
N.K. also thanks the organizers of ``SUSY17" for providing 
the opportunity to present and discuss the results of this paper.

%=========================
\bibliographystyle{JHEP}
\bibliography{collider}

\providecommand{\href}[2]{#2}\begingroup\raggedright\begin{thebibliography}{10}

\bibitem{Ko:2015uma}
P.~Ko and T.~Nomura, {\it {SU(2)$_L\times$SU(2)$_R$ minimal dark matter with 2
  TeV $W'$}},  {\em Phys. Lett.} {\bf B753} (2016) 612--618,
  [\href{http://arxiv.org/abs/1510.07872}{{\tt arXiv:1510.07872}}].

\bibitem{Agarwalla:2016rmw}
S.~K. Agarwalla, K.~Ghosh, and A.~Patra, {\it {LHC diphoton excess in a
  left-right symmetric model with minimal dark matter}},
  \href{http://arxiv.org/abs/1607.03878}{{\tt arXiv:1607.03878}}.

\bibitem{Agarwalla:2018lnq}
S.~Kumar~Agarwalla, K.~Ghosh, and A.~Patra, {\it {Sub-TeV Quintuplet Minimal
  Dark Matter with Left-Right Symmetry}},  {\em JHEP} {\bf 05} (2018) 123,
  [\href{http://arxiv.org/abs/1803.01670}{{\tt arXiv:1803.01670}}].

\bibitem{Cirelli:2005uq}
M.~Cirelli, N.~Fornengo, and A.~Strumia, {\it {Minimal dark matter}},  {\em
  Nucl. Phys.} {\bf B753} (2006) 178--194,
  [\href{http://arxiv.org/abs/hep-ph/0512090}{{\tt hep-ph/0512090}}].

\bibitem{Heeck:2015qra}
J.~Heeck and S.~Patra, {\it {Minimal Left-Right Symmetric Dark Matter}},  {\em
  Phys. Rev. Lett.} {\bf 115} (2015), no.~12 121804,
  [\href{http://arxiv.org/abs/1507.01584}{{\tt arXiv:1507.01584}}].

\bibitem{Garcia-Cely:2015quu}
C.~Garcia-Cely and J.~Heeck, {\it {Phenomenology of left-right symmetric dark
  matter}},  \href{http://arxiv.org/abs/1512.03332}{{\tt arXiv:1512.03332}}.
  [JCAP1603,021(2016)].

\bibitem{Maru:2017pwl}
N.~Maru, N.~Okada, and S.~Okada, {\it {Fermionic Minimal Dark Matter in 5D
  Gauge-Higgs Unification}},  {\em Phys. Rev.} {\bf D96} (2017), no.~11 115023,
  [\href{http://arxiv.org/abs/1801.00686}{{\tt arXiv:1801.00686}}].

\bibitem{Ostdiek:2015aga}
B.~Ostdiek, {\it {Constraining the minimal dark matter fiveplet with LHC
  searches}},  {\em Phys. Rev.} {\bf D92} (2015) 055008,
  [\href{http://arxiv.org/abs/1506.03445}{{\tt arXiv:1506.03445}}].

\bibitem{Kumericki:2012bh}
K.~Kumericki, I.~Picek, and B.~Radovcic, {\it {TeV-scale Seesaw with Quintuplet
  Fermions}},  {\em Phys. Rev.} {\bf D86} (2012) 013006,
  [\href{http://arxiv.org/abs/1204.6599}{{\tt arXiv:1204.6599}}].

\bibitem{Yu:2015pwa}
Y.~Yu, C.-X. Yue, and S.~Yang, {\it {Signatures of the quintuplet leptons at
  the LHC}},  {\em Phys. Rev.} {\bf D91} (2015), no.~9 093003,
  [\href{http://arxiv.org/abs/1502.02801}{{\tt arXiv:1502.02801}}].

\bibitem{Mohapatra:1974hk}
R.~N. Mohapatra and J.~C. Pati, {\it {Left-Right Gauge Symmetry and an
  Isoconjugate Model of CP Violation}},  {\em Phys. Rev.} {\bf D11} (1975)
  566--571.

\bibitem{Senjanovic:1975rk}
G.~Senjanovic and R.~N. Mohapatra, {\it {Exact Left-Right Symmetry and
  Spontaneous Violation of Parity}},  {\em Phys. Rev.} {\bf D12} (1975) 1502.

\bibitem{Beg:1978mt}
M.~A.~B. Beg and H.~S. Tsao, {\it {Strong P, T Noninvariances in a Superweak
  Theory}},  {\em Phys. Rev. Lett.} {\bf 41} (1978) 278.

\bibitem{Mohapatra:1978fy}
R.~N. Mohapatra and G.~Senjanovic, {\it {Natural Suppression of Strong p and t
  Noninvariance}},  {\em Phys. Lett.} {\bf 79B} (1978) 283--286.

\bibitem{Babu:1989rb}
K.~S. Babu and R.~N. Mohapatra, {\it {A Solution to the Strong {CP} Problem
  Without an Axion}},  {\em Phys. Rev.} {\bf D41} (1990) 1286.

\bibitem{Barr:1991qx}
S.~M. Barr, D.~Chang, and G.~Senjanovic, {\it {Strong CP problem and parity}},
  {\em Phys. Rev. Lett.} {\bf 67} (1991) 2765--2768.

\bibitem{Mohapatra:1995xd}
R.~N. Mohapatra and A.~Rasin, {\it {Simple supersymmetric solution to the
  strong CP problem}},  {\em Phys. Rev. Lett.} {\bf 76} (1996) 3490--3493,
  [\href{http://arxiv.org/abs/hep-ph/9511391}{{\tt hep-ph/9511391}}].

\bibitem{Kuchimanchi:1995rp}
R.~Kuchimanchi, {\it {Solution to the strong CP problem: Supersymmetry with
  parity}},  {\em Phys. Rev. Lett.} {\bf 76} (1996) 3486--3489,
  [\href{http://arxiv.org/abs/hep-ph/9511376}{{\tt hep-ph/9511376}}].

\bibitem{Mohapatra:1997su}
R.~N. Mohapatra, A.~Rasin, and G.~Senjanovic, {\it {P, C and strong CP in
  left-right supersymmetric models}},  {\em Phys. Rev. Lett.} {\bf 79} (1997)
  4744--4747, [\href{http://arxiv.org/abs/hep-ph/9707281}{{\tt
  hep-ph/9707281}}].

\bibitem{Babu:2001se}
K.~S. Babu, B.~Dutta, and R.~N. Mohapatra, {\it {Solving the strong CP and the
  SUSY phase problems with parity symmetry}},  {\em Phys. Rev.} {\bf D65}
  (2002) 016005, [\href{http://arxiv.org/abs/hep-ph/0107100}{{\tt
  hep-ph/0107100}}].

\bibitem{Kuchimanchi:2010xs}
R.~Kuchimanchi, {\it {P/CP Conserving CP/P Violation Solves Strong CP
  Problem}},  {\em Phys. Rev.} {\bf D82} (2010) 116008,
  [\href{http://arxiv.org/abs/1009.5961}{{\tt arXiv:1009.5961}}].

\bibitem{Peccei:1977hh}
R.~D. Peccei and H.~R. Quinn, {\it {CP Conservation in the Presence of
  Instantons}},  {\em Phys. Rev. Lett.} {\bf 38} (1977) 1440--1443.

\bibitem{Mukhopadhyaya:2010qf}
B.~Mukhopadhyaya and S.~Mukhopadhyay, {\it {Same-sign trileptons and
  four-leptons as signatures of new physics at the CERN Large Hadron
  Collider}},  {\em Phys. Rev.} {\bf D82} (2010) 031501,
  [\href{http://arxiv.org/abs/1005.3051}{{\tt arXiv:1005.3051}}].

\bibitem{Mukhopadhyay:2011xs}
S.~Mukhopadhyay and B.~Mukhopadhyaya, {\it {Same-sign trileptons at the LHC: A
  Window to lepton-number violating supersymmetry}},  {\em Phys. Rev.} {\bf
  D84} (2011) 095001, [\href{http://arxiv.org/abs/1108.4921}{{\tt
  arXiv:1108.4921}}].

\bibitem{Bambhaniya:2013yca}
G.~Bambhaniya, J.~Chakrabortty, S.~Goswami, and P.~Konar, {\it {Generation of
  neutrino mass from new physics at TeV scale and multilepton signatures at the
  LHC}},  {\em Phys. Rev.} {\bf D88} (2013), no.~7 075006,
  [\href{http://arxiv.org/abs/1305.2795}{{\tt arXiv:1305.2795}}].

\bibitem{Chun:2012zu}
E.~J. Chun and P.~Sharma, {\it {Same-Sign Tetra-Leptons from Type II Seesaw}},
  {\em JHEP} {\bf 08} (2012) 162, [\href{http://arxiv.org/abs/1206.6278}{{\tt
  arXiv:1206.6278}}].

\bibitem{Sirunyan:2017uyt}
{\bf CMS} Collaboration, A.~M. Sirunyan et~al., {\it {Search for physics beyond
  the standard model in events with two leptons of same sign, missing
  transverse momentum, and jets in proton--proton collisions at $\sqrt{s} =
  13\,\text {TeV} $}},  {\em Eur. Phys. J.} {\bf C77} (2017), no.~9 578,
  [\href{http://arxiv.org/abs/1704.07323}{{\tt arXiv:1704.07323}}].

\bibitem{Aad:2011vj}
{\bf ATLAS} Collaboration, G.~Aad et~al., {\it {Inclusive search for same-sign
  dilepton signatures in $pp$ collisions at $\sqrt{s}=7$ TeV with the ATLAS
  detector}},  {\em JHEP} {\bf 10} (2011) 107,
  [\href{http://arxiv.org/abs/1108.0366}{{\tt arXiv:1108.0366}}].

\bibitem{Aad:2015mia}
{\bf ATLAS} Collaboration, G.~Aad et~al., {\it {Search for squarks and gluinos
  in events with isolated leptons, jets and missing transverse momentum at
  $\sqrt{s}=8$ TeV with the ATLAS detector}},  {\em JHEP} {\bf 04} (2015) 116,
  [\href{http://arxiv.org/abs/1501.03555}{{\tt arXiv:1501.03555}}].

\bibitem{Aaboud:2017bac}
{\bf ATLAS} Collaboration, M.~Aaboud et~al., {\it {Search for squarks and
  gluinos in events with an isolated lepton, jets, and missing transverse
  momentum at $\sqrt{s}=13$ TeV with the ATLAS detector}},  {\em Phys. Rev.}
  {\bf D96} (2017), no.~11 112010, [\href{http://arxiv.org/abs/1708.08232}{{\tt
  arXiv:1708.08232}}].

\bibitem{Aaboud:2018ujj}
{\bf ATLAS} Collaboration, M.~Aaboud et~al., {\it {Search for new phenomena
  using the invariant mass distribution of same-flavour opposite-sign dilepton
  pairs in events with missing transverse momentum in $\sqrt{s}=13$ $\text
  {Te}\text {V}$ pp collisions with the ATLAS detector}},  {\em Eur. Phys. J.}
  {\bf C78} (2018), no.~8 625, [\href{http://arxiv.org/abs/1805.11381}{{\tt
  arXiv:1805.11381}}].

\bibitem{Sirunyan:2018iwl}
{\bf CMS} Collaboration, A.~M. Sirunyan et~al., {\it {Search for new physics in
  events with two soft oppositely charged leptons and missing transverse
  momentum in proton-proton collisions at $\sqrt{s}=$ 13 TeV}},
  \href{http://arxiv.org/abs/1801.01846}{{\tt arXiv:1801.01846}}.

\bibitem{Minkowski:1977sc}
P.~Minkowski, {\it {$\mu \to e\gamma$ at a Rate of One Out of $10^{9}$ Muon
  Decays?}},  {\em Phys. Lett.} {\bf 67B} (1977) 421--428.

\bibitem{Yanagida:1979as}
T.~Yanagida, {\it {HORIZONTAL SYMMETRY AND MASSES OF NEUTRINOS}},  {\em Conf.
  Proc.} {\bf C7902131} (1979) 95--99.

\bibitem{Sawada:1979dis}
O.~Sawada and A.~Sugamoto, eds., {\em {Proceedings: Workshop on the Unified
  Theories and the Baryon Number in the Universe}}, (Tsukuba, Japan),
  Natl.Lab.High Energy Phys., Natl.Lab.High Energy Phys., 1979.

\bibitem{Levy:1980ws}
M.~Levy, J.~L. Basdevant, D.~Speiser, J.~Weyers, R.~Gastmans, and M.~Jacob,
  eds., {\em {QUARKS AND LEPTONS. PROCEEDINGS, SUMMER INSTITUTE, CARGESE,
  FRANCE, JULY 9-29, 1979}}, vol.~61, 1980.

\bibitem{VanNieuwenhuizen:1979hm}
P.~Van~Nieuwenhuizen and D.~Z. Freedman, eds., {\em {SUPERGRAVITY. PROCEEDINGS,
  WORKSHOP AT STONY BROOK, 27-29 SEPTEMBER 1979}}, 1979.

\bibitem{Mohapatra:1979ia}
R.~N. Mohapatra and G.~Senjanovic, {\it {Neutrino Mass and Spontaneous Parity
  Violation}},  {\em Phys. Rev. Lett.} {\bf 44} (1980) 912.

\bibitem{Chang:1983fu}
D.~Chang, R.~N. Mohapatra, and M.~K. Parida, {\it {Decoupling Parity and
  SU(2)-R Breaking Scales: A New Approach to Left-Right Symmetric Models}},
  {\em Phys. Rev. Lett.} {\bf 52} (1984) 1072.

\bibitem{Ball:2014uwa}
{\bf NNPDF} Collaboration, R.~D. Ball et~al., {\it {Parton distributions for
  the LHC Run II}},  {\em JHEP} {\bf 04} (2015) 040,
  [\href{http://arxiv.org/abs/1410.8849}{{\tt arXiv:1410.8849}}].

\bibitem{Ball:2013hta}
{\bf NNPDF} Collaboration, R.~D. Ball, V.~Bertone, S.~Carrazza, L.~Del~Debbio,
  S.~Forte, A.~Guffanti, N.~P. Hartland, and J.~Rojo, {\it {Parton
  distributions with QED corrections}},  {\em Nucl. Phys.} {\bf B877} (2013)
  290--320, [\href{http://arxiv.org/abs/1308.0598}{{\tt arXiv:1308.0598}}].

\bibitem{Martin:2004dh}
A.~D. Martin, R.~G. Roberts, W.~J. Stirling, and R.~S. Thorne, {\it {Parton
  distributions incorporating QED contributions}},  {\em Eur. Phys. J.} {\bf
  C39} (2005) 155--161, [\href{http://arxiv.org/abs/hep-ph/0411040}{{\tt
  hep-ph/0411040}}].

\bibitem{Schmidt:2015zda}
C.~Schmidt, J.~Pumplin, D.~Stump, and C.~P. Yuan, {\it {CT14QED parton
  distribution functions from isolated photon production in deep inelastic
  scattering}},  {\em Phys. Rev.} {\bf D93} (2016), no.~11 114015,
  [\href{http://arxiv.org/abs/1509.02905}{{\tt arXiv:1509.02905}}].

\bibitem{Alloul:2013bka}
A.~Alloul, N.~D. Christensen, C.~Degrande, C.~Duhr, and B.~Fuks, {\it
  {FeynRules 2.0 - A complete toolbox for tree-level phenomenology}},  {\em
  Comput. Phys. Commun.} {\bf 185} (2014) 2250--2300,
  [\href{http://arxiv.org/abs/1310.1921}{{\tt arXiv:1310.1921}}].

\bibitem{Christensen:2008py}
N.~D. Christensen and C.~Duhr, {\it {FeynRules - Feynman rules made easy}},
  {\em Comput. Phys. Commun.} {\bf 180} (2009) 1614--1641,
  [\href{http://arxiv.org/abs/0806.4194}{{\tt arXiv:0806.4194}}].

\bibitem{Alwall:2014hca}
J.~Alwall, R.~Frederix, S.~Frixione, V.~Hirschi, F.~Maltoni, O.~Mattelaer,
  H.~S. Shao, T.~Stelzer, P.~Torrielli, and M.~Zaro, {\it {The automated
  computation of tree-level and next-to-leading order differential cross
  sections, and their matching to parton shower simulations}},  {\em JHEP} {\bf
  07} (2014) 079, [\href{http://arxiv.org/abs/1405.0301}{{\tt
  arXiv:1405.0301}}].

\bibitem{Ball:2012cx}
R.~D. Ball et~al., {\it {Parton distributions with LHC data}},  {\em Nucl.
  Phys.} {\bf B867} (2013) 244--289,
  [\href{http://arxiv.org/abs/1207.1303}{{\tt arXiv:1207.1303}}].

\bibitem{Alwall:2011uj}
J.~Alwall, M.~Herquet, F.~Maltoni, O.~Mattelaer, and T.~Stelzer, {\it {MadGraph
  5 : Going Beyond}},  {\em JHEP} {\bf 06} (2011) 128,
  [\href{http://arxiv.org/abs/1106.0522}{{\tt arXiv:1106.0522}}].

\bibitem{Sjostrand:2006za}
T.~Sjostrand, S.~Mrenna, and P.~Z. Skands, {\it {PYTHIA 6.4 Physics and
  Manual}},  {\em JHEP} {\bf 05} (2006) 026,
  [\href{http://arxiv.org/abs/hep-ph/0603175}{{\tt hep-ph/0603175}}].

\bibitem{deFavereau:2013fsa}
{\bf DELPHES 3} Collaboration, J.~de~Favereau, C.~Delaere, P.~Demin,
  A.~Giammanco, V.~Lema{\^\i}tre, A.~Mertens, and M.~Selvaggi, {\it {DELPHES 3,
  A modular framework for fast simulation of a generic collider experiment}},
  {\em JHEP} {\bf 02} (2014) 057, [\href{http://arxiv.org/abs/1307.6346}{{\tt
  arXiv:1307.6346}}].

\bibitem{Khachatryan:2014sta}
{\bf CMS} Collaboration, V.~Khachatryan et~al., {\it {Study of vector boson
  scattering and search for new physics in events with two same-sign leptons
  and two jets}},  {\em Phys. Rev. Lett.} {\bf 114} (2015), no.~5 051801,
  [\href{http://arxiv.org/abs/1410.6315}{{\tt arXiv:1410.6315}}].

\bibitem{Campbell:1999ah}
J.~M. Campbell and R.~K. Ellis, {\it {An Update on vector boson pair production
  at hadron colliders}},  {\em Phys. Rev.} {\bf D60} (1999) 113006,
  [\href{http://arxiv.org/abs/hep-ph/9905386}{{\tt hep-ph/9905386}}].

\bibitem{Campanario:2008yg}
F.~Campanario, V.~Hankele, C.~Oleari, S.~Prestel, and D.~Zeppenfeld, {\it {QCD
  corrections to charged triple vector boson production with leptonic decay}},
  {\em Phys. Rev.} {\bf D78} (2008) 094012,
  [\href{http://arxiv.org/abs/0809.0790}{{\tt arXiv:0809.0790}}].

\bibitem{Garzelli:2012bn}
M.~V. Garzelli, A.~Kardos, C.~G. Papadopoulos, and Z.~Trocsanyi, {\it {t
  $\bar{t}$ $W^{+-}$ and t $\bar{t}$ Z Hadroproduction at NLO accuracy in QCD
  with Parton Shower and Hadronization effects}},  {\em JHEP} {\bf 11} (2012)
  056, [\href{http://arxiv.org/abs/1208.2665}{{\tt arXiv:1208.2665}}].

\bibitem{Heinemeyer:2013tqa}
{\bf LHC Higgs Cross Section Working Group} Collaboration, J.~R. Andersen
  et~al., {\it {Handbook of LHC Higgs Cross Sections: 3. Higgs Properties}},
  \href{http://arxiv.org/abs/1307.1347}{{\tt arXiv:1307.1347}}.

\bibitem{Yong-Bai:2015xna}
Y.-B. Shen, R.-Y. Zhang, W.-G. Ma, X.-Z. Li, Y.~Zhang, and L.~Guo, {\it {NLO
  QCD + NLO EW corrections to $WZZ$ productions with leptonic decays at the
  LHC}},  {\em JHEP} {\bf 10} (2015) 186,
  [\href{http://arxiv.org/abs/1507.03693}{{\tt arXiv:1507.03693}}]. [Erratum:
  JHEP10,156(2016)].

\bibitem{Nhung:2013jta}
D.~T. Nhung, L.~D. Ninh, and M.~M. Weber, {\it {NLO corrections to WWZ
  production at the LHC}},  {\em JHEP} {\bf 12} (2013) 096,
  [\href{http://arxiv.org/abs/1307.7403}{{\tt arXiv:1307.7403}}].

\bibitem{Bredenstein:2009aj}
A.~Bredenstein, A.~Denner, S.~Dittmaier, and S.~Pozzorini, {\it {NLO QCD
  corrections to pp $->$ t anti-t b anti-b + X at the LHC}},  {\em Phys. Rev.
  Lett.} {\bf 103} (2009) 012002, [\href{http://arxiv.org/abs/0905.0110}{{\tt
  arXiv:0905.0110}}].

\bibitem{Bevilacqua:2012em}
G.~Bevilacqua and M.~Worek, {\it {Constraining BSM Physics at the LHC: Four top
  final states with NLO accuracy in perturbative QCD}},  {\em JHEP} {\bf 07}
  (2012) 111, [\href{http://arxiv.org/abs/1206.3064}{{\tt arXiv:1206.3064}}].

\bibitem{Cowan:SLAC2012}
G.~Cowan, {\it Two developments in discovery tests: use of weighted monte carlo
  events and an improved measure of experimental sensitivity}, . Talk given
  during the meeting on Progress on Statistical Issues in Searches, June 4-6,
  2012, SLAC, USA, {\tt
  http://www-conf.slac.stanford.edu/statisticalissues2012/talks/glen\_cowan\_slac\_4jun12.pdf}.

\bibitem{Kumar:2015tna}
N.~Kumar and S.~P. Martin, {\it {Vectorlike leptons at the Large Hadron
  Collider}},  {\em Phys. Rev.} {\bf D92} (2015), no.~11 115018,
  [\href{http://arxiv.org/abs/1510.03456}{{\tt arXiv:1510.03456}}].

\bibitem{Ozcan:2009qm}
V.~E. Ozcan, S.~Sultansoy, and G.~Unel, {\it {Possible Discovery Channel for
  New Charged Leptons at the LHC}},  {\em J. Phys.} {\bf G36} (2009) 095002,
  [\href{http://arxiv.org/abs/0903.3177}{{\tt arXiv:0903.3177}}]. [Erratum: J.
  Phys.G37,059801(2010)].

\end{thebibliography}\endgroup
%==========================

\end{document}